\documentclass[pra,aps]{revtex4-2}
\usepackage{amsmath}
\usepackage{amssymb}
\usepackage{graphicx}
\usepackage{dcolumn}
\usepackage{bm}

\usepackage{multirow} 
\usepackage{colortbl}  
\definecolor{DGray}{gray}{0.85}
\definecolor{Gray}{gray}{0.95}
\usepackage{diagbox} 
\usepackage{siunitx}
\usepackage{subcaption}
\usepackage{caption}
\begin{document}


\title{Sensitivity of an antineutrino monitor for remote nuclear reactor discovery}


\author{L. Kneale}
\email[Corresponding author (she/her/hers): ]{e.kneale@sheffield.ac.uk}
\author{S. T. Wilson}%
\email[Corresponding author (he/him/his): ]{stephen.wilson@sheffield.ac.uk}
\author{T. Appleyard} 
\author{J. Armitage}
\author{N. Holland}
\author{M. Malek}
\affiliation{%
University of Sheffield, Sheffield S10 2TN, United Kingdom
}%

\date{\today}

\begin{abstract}
Antineutrinos from a nuclear reactor comprise an unshieldable signal which carries information about the core. A gadolinium-doped, water-based Cherenkov detector could detect reactor antineutrinos for mid- to far-field remote reactor monitoring for non-proliferation applications.

Two novel and independent reconstruction and analysis pathways have been developed and applied to a number of representative reactor signals to evaluate the sensitivity of a kiloton-scale, gadolinium-doped Cherenkov detector as a remote monitor. 

The sensitivity of four detector configurations to nine reactor signal combinations was evaluated for a detector situated in Boulby Mine, close to the Boulby Underground Laboratory in the UK. It was found that a 22~m detector with a gadolinium-doped, water-based liquid scintillator fill is sensitive to a $\rm \sim 3~GW_{th}$ reactor at a standoff of $\rm \sim150~km$ within two years in the current reactor landscape. A larger detector would be required to achieve a more timely detection or to monitor smaller or more distant reactors.
\end{abstract}

\keywords{Reactor antineutrinos, gadolinium, reactor monitoring, water based liquid scintillator, water Cherenkov}
\maketitle

\section{Introduction}

Emerging antineutrino detection technology has opened up the possibility of using a water-based detector as a remote monitoring tool for nuclear non-proliferation. While near-field reactor observation with surface-deployed, plastic scintillator detectors has been demonstrated~\cite{Kuroda2012,Haghighat2020} and recently investigated with a view to planned reactor monitoring~\cite{Ozturk2020}, an underground water-based detector allows the technology to be scaled to larger detector sizes for mid- to far-field monitoring. Aggregate detection of reactor antineutrinos has been achieved for the first time in pure water~\cite{SNO2022} and applications of antineutrino detection to nuclear non-proliferation have been explored in~\cite{Bernstein2021}. A remote monitor of this type may be used in a collaborative project as part of a remote monitoring toolkit for verification of declared reactor activity. There is interest in the safeguarding and policy communities in a neutrino detector as a future tool to safeguard advanced reactors and as part of future nuclear deals~\cite{NuTools}, with a reduction in the perceived intrusion of on-site inspections at nuclear facilities cited as one of the key benefits. Although the high cost, large size and underground location of large water-based neutrino detectors limit their practical application for non-cooperative reactor monitoring, these concerns could be somewhat mitigated by adoption of a neutrino monitor at an early stage in the construction of an advanced reactor or development of a nuclear deal. 

This paper provides a first attempt to evaluate the sensitivity of a neutrino detector to real reactor signals, and analytical and statistical methods that could be employed to draw conclusions about the operation of a reactor in a realistic nuclear landscape.

The challenge of remote reactor monitoring with a water-based detector is that the low-energy antineutrinos from a reactor are at the very limit of the energy threshold. Backgrounds (\textit{e.g.}, from other reactors and natural radioactivity of detector components) further reduce sensitivity to the signal. This issue requires a multi-faceted approach to improving the sensitivity. This paper presents two new complementary reconstruction-analysis pathways for signal-to-background discrimination, which have been developed for optimal sensitivity to a remote reactor. The methods presented in this paper represent a comprehensive treatment of reactor antineutrino detection to evaluate sensitivity to nine real reactor signals with four different detector configurations close to the Science \& Technology Facilities Council (STFC) Boulby Underground Laboratory in the UK.

The paper is structured as follows. Section~\ref{sec:reactor-antinu} presents the fundamentals of reactor antineutrino emission and detection in water-based media. The site at Boulby is detailed in Section~\ref{sec:boulby-detector}, along with discussion of the detector configurations and signals evaluated. In Section~\ref{sec:sims}, shared signal and background simulations are detailed and the reconstruction-analysis pathways are discussed in depth in Section~\ref{sec:sensitivity}. Results for the sensitivity of detector-signal configurations are presented and discussed in Section~\ref{sec:results} before concluding in Section~\ref{sec:Conclusions}. 

\section{Reactor Antineutrinos}
\label{sec:reactor-antinu}

All nuclear power reactors generate antineutrinos, emitting an isotropic flux of $\rm \mathcal{O}(10^{20})~s^{-1}$ from a $\rm 1 ~GW_{th}$ reactor~\cite{Porta2010a}. This is produced by the fission of $\rm ^{235}$U, $\rm ^{238}$U, $\rm ^{2
39}$Pu and $\rm^{241}$Pu into neutron-rich nuclei, which then undergo a series of $\rm \beta$ decays to stability. This releases, on average, 6~antineutrinos per fission at energies up to $\rm \sim10$~MeV. Although the interaction cross section of antineutrinos with matter is very small, the enormous number of antineutrinos released means that the antineutrino signal from a reactor can be seen in a variety of detectors.

The reactor antineutrino flux is dependent on the reactor thermal power and core composition, on the nuclear physics of the fission of the isotopes in the core and on neutrino oscillations, which alter the flux with distance from a reactor. The precise composition of the core of a reactor and the time evolution of the core (\textit{burnup}), including the refueling frequency, depend on the reactor type.

The calculation of the antineutrino spectrum from a reactor is described in~\cite{Dye2021}. The antineutrino spectrum from a fissioning isotope in a reactor is related to the power output and composition of a reactor core by 

\begin{equation}
    \Phi_{\Bar{\nu}_e,i}(E_{\Bar{\nu}_e}) = P_{th} \frac{p_\textit{i} \lambda_{\textit{i}}(E_{\Bar{\nu}_e})}{Q_{\textit{i}}},
    \label{eq:emitted_flux}
\end{equation}
where $\mathit{P_{th}}$ is the thermal power of the core, $\mathit{p_i}$ is the fraction of the thermal power resulting from the fission of isotope $\textit{i}$, $\mathit{Q_i}$ is the average thermal energy emitted per fission and $\lambda_{\textit{i}}(E_{\Bar{\nu}_e})$ is the emission energy spectrum in antineutrinos per fission for fissioning isotope \textit{i} as a function of antineutrino energy $E_{\Bar{\nu}_e}$ given by
\begin{equation}
    \lambda_i(E_{\Bar{\nu}_e}) = exp \Bigg( \sum\limits_{j=1}^6 a_j E^{j-1}_{\Bar{\nu}_e}\Bigg),
    \label{equation:spectrum}
\end{equation}
where the coefficients $\mathit{a_j}$ are fit parameters from the Huber-Mueller predictions~\cite{Huber2011,Mueller2011}, which are derived from measurements of the $\beta$ spectra from nuclear fission. 

Information about the power and core composition of a reactor is carried by the outgoing particles from antineutrino interactions in a detector. A change in the number of antineutrinos emitted by a reactor can be caused by a change in the core composition or reactor thermal power, or indeed by both. The SONGS1 gadolinium-doped liquid-scintillator antineutrino detector, located 25~m from the $\rm 3.56~GW_{th}$ San Onofre Nuclear Generating Station (SONGS), demonstrated that the detected antineutrino flux from a reactor reflects the operating power and fuel evolution of the core~\cite{Bowden2009} and as such could be used for discovery, monitoring and verification of nuclear reactor operations. 

\subsection{Reactor antineutrino detection}

Antineutrinos from a reactor can be detected via their inverse $\rm \beta$ decay (IBD) interaction with protons in water or a hydrogenated liquid: 
$$\rm \overline{\nu}_e + p \longrightarrow e^+ + n.$$ 

Information about the energy of the incoming antineutrino is carried by the outgoing positron, while neutron tagging can help to reject backgrounds and lower the detectable energy threshold. IBD is the principle interaction by which antineutrinos can be detected for reactor monitoring in a water-based Cherenkov detector.

The IBD cross section is $ \mathcal{O} (10^{-44})E_e p_e~\rm{cm^2}$~\cite{Vogel1999}, where $E_e$ and $p_e$ are the positron energy and momentum. Although small, it is relatively high compared to the cross sections of other antineutrino interactions in matter and it has been calculated to within 1\% accuracy at low energies. At the time of this study, the most accurate cross section in the MeV to GeV range was given in~\cite{Strumia2003}. The cross section has since been recalculated with reduced uncertainty in~\cite{Ricciardi2022}.

IBD is the dominant interaction of antineutrinos with energies of less than a few tens of MeV and has a low threshold energy $\mathit{E_{thr}}$ which can be expressed in terms of the proton, neutron and positron rest masses $\mathit{m_p}$, $\mathit{m_n}$ and $\mathit{m_e}$ as approximately

\begin{equation}
    E_{thr} \approx \frac{(m_n + m_e)^2 - m_p^2}{2m_p} \approx 1.8 \rm~MeV
\end{equation} 
in the laboratory frame.

The positron carries most of the kinetic energy from the antineutrino and, with good energy resolution, the incident antineutrino energy can be determined. The energy of the incoming antineutrino $E_{\Bar{\nu}_e}$ is related approximately to that of the positron by
\begin{equation}
    E_{\Bar{\nu}_e} \approx E_{e^+}  + E_{thr} - m_e.
    \label{equation:positronenergy}
\end{equation}

While the positron emission is almost isotropic, with a slight bias in the backwards direction, the neutron takes on most of the antineutrino's momentum and its initial direction is largely parallel to that of the incoming antineutrino. From the point of emission, the neutron then takes a random walk and thermalizes in the detector medium through successive scatterings, which knock the neutron off its original path. Once thermalized, the neutron is captured on a hydrogen nucleus or on another nucleus, such as gadolinium, added specifically for its neutron-capture capabilities. A second signal arising from the de-excitation of the capture nucleus can be detected. This occurs within a short distance and time of the positron signal and results in a signal of coincident interactions which can be beneficial for background rejection. The time and distance between the positron and neutron events are dependent on the medium in which the interaction takes place.

\subsection{Reactor antineutrino detection media}

Water Cherenkov and scintillator detectors are the two principal types of antineutrino detection technology. Scintillator detectors are a proven technology for reactor antineutrino detection but are not readily scalable for mid- to far-field monitoring. A nascent water Cherenkov technology - gadolinium doping - presents the possibility of a scalable reactor antineutrino detector. The combination of the two technologies into a water-based scintillator technology promises to exploit the best features of each method.

The principle of using gadolinium (Gd) to delve into lower-energy neutrino detection was first introduced by~\cite{Bernstein2001} and developed by~\cite{Beacom2003}. Gadolinium has a very high thermal neutron capture cross section ($48,800$ barns (b) for natural Gd compared to $\sim0.3$~b for  hydrogen) and a relatively high-energy subsequent gamma cascade of $\sim8$~MeV (mean total energy) compared to a single 2.2~MeV gamma from the capture on hydrogen. This gives a more easily detectable correlated signal from the inverse $\beta$ decay reaction~\cite{Beacom2003}. With a concentration of 0.1\% Gd ions, $\sim90$\% of the neutrons capture on Gd~\cite{Beacom2003}. Most of the remaining neutrons capture onto the hydrogen in the water.
  
In 0.1\% gadolinium-doped ultra-pure water (Gd-H$_2$O), the neutron thermalizes and captures after a mean time of $\rm \sim30~\rm \mu s$ and mean distance of $\sim 6$~cm. The delayed neutron-capture emission is seen in a Gd-H$_2$O Cherenkov detector with a peak in visible light at $\sim4.5$~MeV. The peak positron energy from IBD interactions of reactor antineutrinos is $\sim2.5$~MeV. In a gadolinium-doped medium, the positron-detection efficiency for reactor antineutrinos can be increased by looking for a positron-like signal (the \textit{prompt} event) in coincidence with the generally higher-energy (and thus easier to observe) neutron-capture signal (the \textit{delayed} event). Positron events at the lower end of the energy range would otherwise be lost among the background. In this way, Gd can lower the energy threshold of a water Cherenkov detector to increase sensitivity to the low-energy positrons for reactor antineutrino interactions via IBD.
  
The emerging Gd-H$_2$O technology has been demonstrated in EGADS (Evaluating Gadolinium's Actions on Detector Systems)~\cite{Marti2020} and has now been deployed in SK-Gd (Super-Kamiokande with Gadolinium)~\cite{ABE2021} and the ANNIE (Accelerator Neutrino-Neutron Interaction Experiment) detector~\cite{back2017}.

Reactor antineutrinos have been detected with liquid and plastic scintillator detectors. A liquid scintillator is composed of an organic solvent containing a scintillating chemical in solution of the type used in the Kamioka Liquid Scintillator Antineutrino Detector (KamLAND)~\cite{Gando2011}. In a scintillating medium, the scintillator interacts with incoming particles, which impart energy to the scintillator. Excited scintillator particles then release this additional energy as light. 

Scintillation detectors bring a high light yield, low-energy sensitivity, and good energy and position resolution. However, they do not preserve directional information and are limited in size due to light absorption and the cost and availability of the medium. Water Cherenkov detectors bring directional information and are scalable to very large detectors. However, they have a low light yield and no sensitivity below the Cherenkov threshold. Combining the two media into a water-based liquid scintillator (WbLS)~\cite{Yeh2011} provides a solution which can be scaled to large sizes and results in a higher light yield, sensitivity down to lower energies, improved energy and position resolution, and directional information from the Cherenkov light, which has benefits for reactor antineutrino detection~\cite{Zsoldos2022}.

WbLS is an emerging detector medium, which is still undergoing optimization and improvement. WbLS cocktails using the PPO (2,5-diphenyl-oxazole) wavelength-shifting scintillator in a linear alkylbenzene (LAB) solvent have been produced~\cite{Yeh2011} and gadolinium doping is in development. Pure liquid scintillator is a scintillating material in solution in an oily organic solvent. In WbLS, the scintillator is dissolved in an oily solvent in the same way. This solution is then further combined with pure water. The mixing between the oil and water in WbLS is achieved by the addition of a surfactant which creates micelles with both hydrophilic and hydrophobic surfaces.

The addition of gadolinium brings a further increase in light yield due to the neutron capture on gadolinium and enhanced background rejection due to the coincident signal pair which occur closer in space and time with gadolinium doping. Gadolinium-doped water-based liquid scintillator (Gd-WbLS) has the potential to combine the benefits of liquid scintillator and gadolinium-doped water Cherenkov detectors. The combination of Cherenkov and scintillation light ultimately brings increased IBD detection, with the added benefit of improved detection quality, particularly where it is possible to separate the Cherenkov and scintillation components.

Although WbLS is an emerging technology - particularly with the addition of Gd - detailed characterization of different WbLS cocktails has been performed~\cite{CHESS,CherenkovScintillation,WbLSTimeResponse,MEV_WbLS_perf} and the ANNIE~\cite{back2017} collaboration is currently working towards a WbLS fill.

\section{Gd-doped Cherenkov detector at Boulby}
\label{sec:boulby-detector}

The location used for this study is the ICL (Israel Chemicals Ltd) Boulby Mine on the coast near Whitby in North Yorkshire, UK. Boulby Mine is an ultra-low background environment and hosts the Boulby Underground Laboratory, which has been home to deep underground physics experiments since the 1990s and is now operated by the UK's STFC.

The low-background environment in Boulby Mine can be attributed to low radioactivity rates in Boulby Mine's rock salt layers, a low cosmic muon rate and the world's lowest ambient air radon concentration in an underground lab ($\rm \sim3 ~Bq ~m^{-3}$). Boulby Underground Laboratory is at a depth of 1.1~km underground and an effective depth of 2.8~km water equivalent (km.w.e.) with a flat overburden. This results in a significant $\mathcal{O}(10^6)$ reduction in cosmogenic muons compared to the surface~\cite{Robinson2003}.

The detector configurations used for this study are based on the AIT-NEO (Advanced Instrumentation Testbed - Neutrino Experiment One) detector, which was investigated for the AIT site at Boulby Mine. In total four configurations - two detector geometries with each of two fill media - and nine reactor signals at Boulby were considered.

\begin{figure}[htb]
    \includegraphics[width=8.6cm]{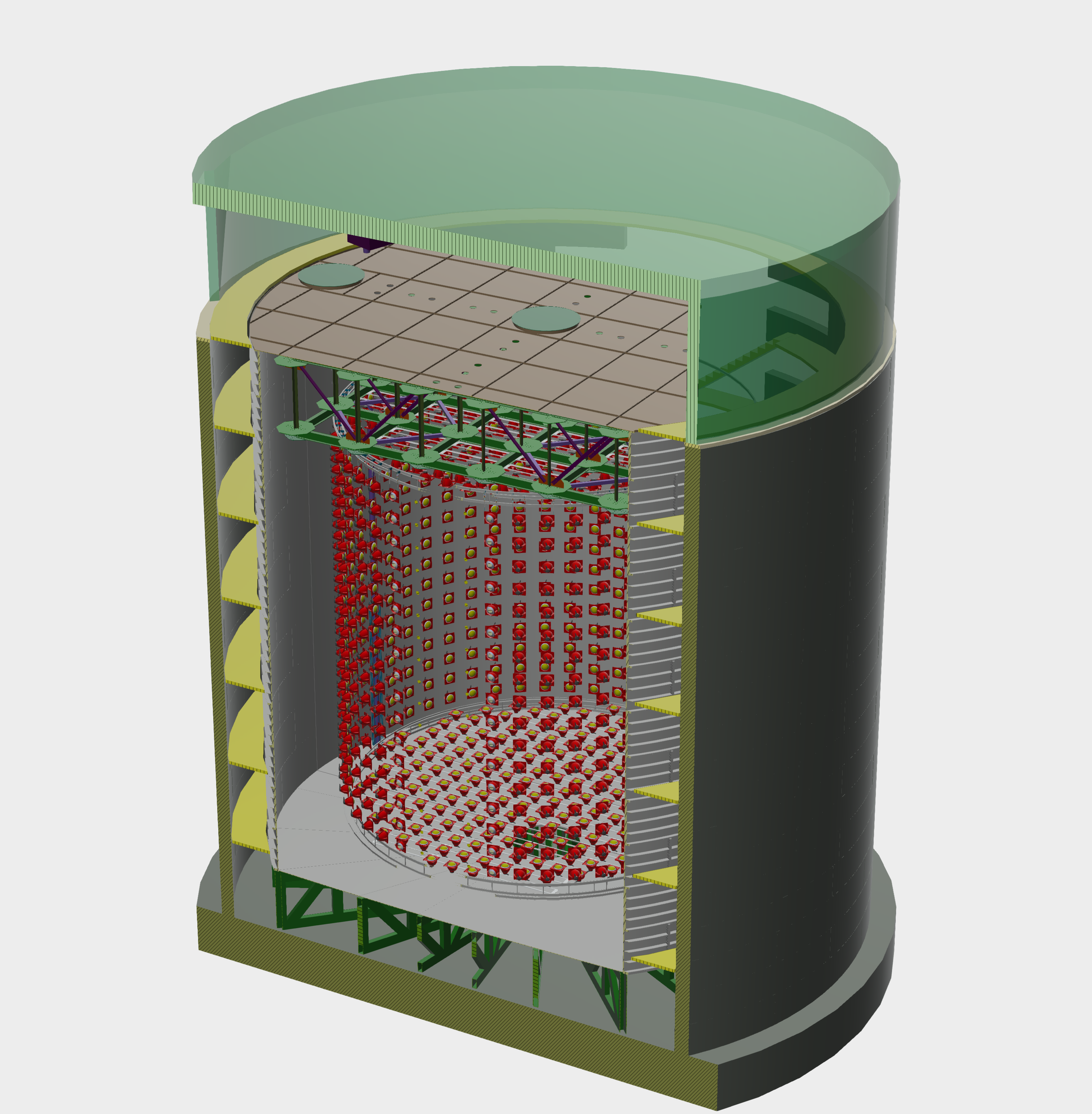}
  \caption{Schematic of the detector design by Jan Boissevain (University of Pennsylvania), showing the tank supported on a steel truss structure and inner PMT support structure.}
  \label{fig:detector-schematic}
\end{figure}

The two detector geometries are upright cylinders, with parameters summarized in Table~\ref{tab:comp-det}. A schematic of the detector design in Fig.~\ref{fig:detector-schematic} shows the inner PMT support structure which creates an instrumented inner detector volume within the tank and an uninstrumented buffer volume between the inner volume and the tank walls for the reduction of backgrounds from the tank and surrounding rock.
\begin{table*}[htb]
    \caption{Summary of detector geometries used in this study.}
\begin{ruledtabular}
 \begin{tabular}{ccccc}
    { Tank diameter}  & { PMT support }         & { Buffer} & { Inner PMT }  & {Number} \\
    { and height [m]} & { structure radius [m]} &    { width [m]} &{ coverage [\%]} & {of PMTs} \\ 
    \hline
    16                   & 5.7                        & 2.3                    & 15  & 2500\\ 
    22                   & 9.0                        & 2.0                      & 15  & 4600\\ 

    \end{tabular}
    \end{ruledtabular}
    \label{tab:comp-det}
\end{table*}

Each of the two detector sizes was evaluated with each of two fill media:
\begin{itemize}
    \item Gd-H$_2$O with 0.2\% Gd$_2$(SO$_4$)$_3$ doping (for 0.1\% Gd concentration) and
    \item Gd-WbLS with 0.2\% Gd$_2$(SO$_4$)$_3$ doping and $\sim100$ photons per MeV WbLS. The WbLS component gives approximately 1\% of the light yield of pure LAB-based scintillator with 2g/L PPO typically used in large neutrino experiments such as Daya Bay and SNO+~\cite{Beriguete2014,Andringa2016}.
\end{itemize}

\begin{figure}[htb]
    \includegraphics[width=8.6cm]{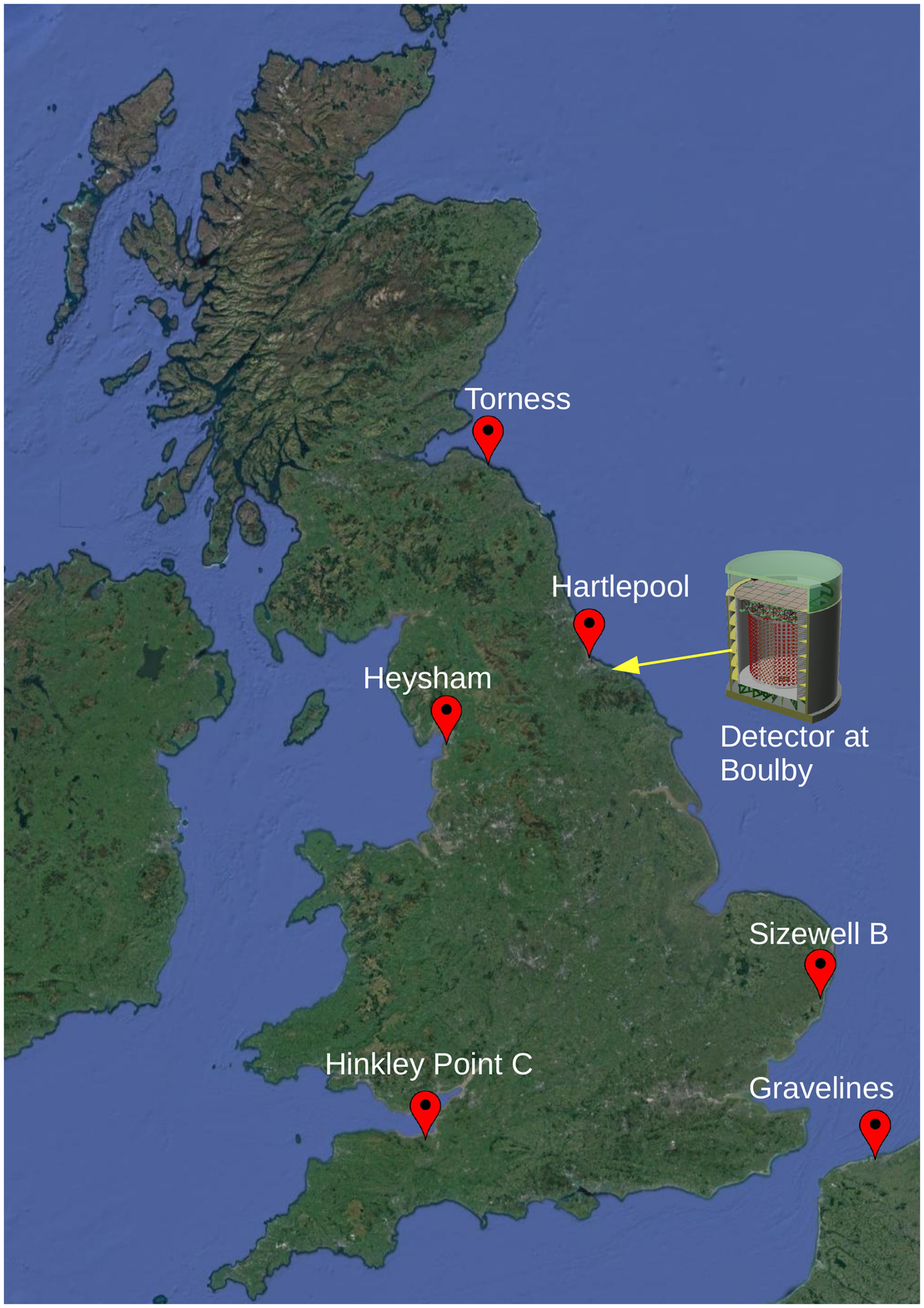}
  \caption[Map]{Map showing the location of the detector at Boulby and the reactor sites studied.~\cite{Google2022}}
  \label{fig:map}
\end{figure}

For this study, the reactor landscape around Boulby Mine is used. Several reactors are planned to decommission or come on line in the coming years and this is reflected in the study. Currently, the principal reactor signals at Boulby are the Advanced Gas-Cooled Reactor (AGR) nuclear power stations at Hartlepool, Heysham and Torness. Other important sources of reactor antineutrinos at Boulby are the Pressurized Water Reactors (PWRs) at Sizewell B and at Gravelines in France. New PWR cores at Hinkley Point C are expected to come on line in the coming years. The detector and reactor locations are shown in Fig.~\ref{fig:map}.

The six principal reactor signals studied are the AGR signals listed in Table~\ref{tab:signal-combos} in order of decreasing signal at Boulby, along with their published dates for decommissioning at the time of this study.

\begin{table*}[htb]
    \centering
        \caption{The six AGR reactor signals evaluated in this study, including their power in terms of total thermal capacity, standoff distance from a detector at Boulby and currently planned decommissioning dates~\cite{EDFEnergy}. Real reactor signals and schedules were used to represent authentic cases of remote monitoring.}
    \begin{ruledtabular}
    \begin{tabular}{lccccc}
    
        {Signal} & {Number} & Reactor &  {Standoff }  & {Decommissioning}  \\
                    &   {of cores}  & power [GW$_{th}$]& {distance [km]} & {date}      \\
        \hline
        Hartlepool 1 \& 2 &  2   &  3.0 &   26     & 2024    \\
        Hartlepool 1      & 1    &  1.5 &   26     & 2024     \\
        Heysham 1 \& 2    & 4    &  3.0 &   149         & 2024 (1), 2028 (2)    \\
        Heysham 2 \& Torness & 4 &  1.5, 3.2 &   149, 187    & 2028        \\
        Heysham 2         & 2    &  1.5 &   149         & 2028      \\
        Torness           & 2    &  3.2 &   187         & 2028       \\
        
    \end{tabular}
    \end{ruledtabular}
    \label{tab:signal-combos}
\end{table*}

Table~\ref{tab:reactor_signals} shows the reactors included in the signal and background for each of the reactor signals listed in Table~\ref{tab:signal-combos}, with the reactor signal and background rates in terms of IBD interactions in a Boulby detector from~\cite{Dye2021}. `World' reactor backgrounds include all reactors more distant than Torness. World reactor backgrounds are projected to 2026, with Hinkley Point B off and Hinkley Point C on, according to published schedules at the time of the study. These scenarios take into account the expected shutdown of the Hartlepool cores and Heysham 1 in 2024.
\begin{table*}
    \centering
    \caption[Combinations of reactor signals and backgrounds evaluated]{Combinations of reactor signals and backgrounds evaluated. Reactors marked black are included in the signal and reactors marked gray are included in the background. Reactors marked with a slash are assumed to be decommissioned. Signal and background rates shown in Neutrino Interaction Units (NIU). NIU (Neutrino Interaction Unit) for IBD is the interaction rate per $10^{32}$ free protons per year. 1 kton $\rm H_2O$ contains $6.686\times10^{31}$ free protons.}
    \begin{ruledtabular}
    \begin{tabular}{lrrrrrr}
    \bf Signal combination &  Hartlepool 1    &  Hartlepool 2     &  Heysham 1       &  Heyhsam 2         &  Torness           &  World \\
    \hline
    \bf Hartlepool 1 \& 2   & \multicolumn{2}{r}{\cellcolor{black}{\textcolor{white}{\bf 1041 }}} & \multicolumn{4}{r}{\cellcolor{DGray}{\bf 207~~}}   \\
    
    \bf Hartlepool 1        & \cellcolor{black}\textcolor{white}{\bf 595} &\multicolumn{5}{r}{\cellcolor{DGray}{ \bf 693~~}} \\
    
    \bf Heysham  1 \& 2      &  \diagbox[innerwidth=5em, height=0.5\line]{}{} &\diagbox[innerwidth=5em, height=0.5\line]{}{}& \multicolumn{2}{r}{\cellcolor{black}{\textcolor{white}{\bf 70}}}   & \multicolumn{2}{r}{\cellcolor{DGray}{\bf 147~~}}   \\
    
       \bf Heysham 2 + Torness   & \diagbox[innerwidth=5em, height=0.5\line]{}{}&\diagbox[innerwidth=5em, height=0.5\line]{}{} &\diagbox[innerwidth=5em, height=0.5\line]{}{} & \multicolumn{2}{r}{\cellcolor{black}{\textcolor{white}{\bf 65}}}  &\cellcolor{DGray} \bf 122~~\\
    
    \bf Heysham 2           &\diagbox[innerwidth=5em, height=0.5\line]{}{} &\diagbox[innerwidth=5em, height=0.5\line]{}{} & \diagbox[innerwidth=5em, height=0.5\line]{}{}& \cellcolor{black} \textcolor{white}{\bf 40}  &  \multicolumn{2}{r}{\cellcolor{DGray} {\bf 147~~}}  \\
    
    \bf Torness             & \diagbox[innerwidth=5em, height=0.5\line]{}{}&\diagbox[innerwidth=5em, height=0.5\line]{}{}  &\diagbox[innerwidth=5em, height=0.5\line]{}{} & \cellcolor{DGray} &  \cellcolor{black} \textcolor{white}{\bf 25} &\cellcolor{DGray}\bf 162\footnote{Total background including Heysham 2 complex.} \\
    \end{tabular}
    \end{ruledtabular}
    \label{tab:reactor_signals}
\end{table*}

An extension of the study to the PWR reactor signals at Sizewell B, Hinkley Point C and Gravelines after 2028 is presented in Section~\ref{subsec:PWRresults}.

\section{Signal and Background Simulations}
\label{sec:sims}

Full Monte Carlo detector simulations were carried out with RAT-PAC (Reactor Analysis Tool - Plus Additional Codes)~\cite{Seibert2014}, which has been adapted for AIT-NEO and which is based on the physics simulation framework GEANT4~\cite{ALLISON2016,AGOSTINELLI2003}, the CLHEP physics library~\cite{LONNBLAD1994}, the GLG4sim (Generic Liquid-scintillator Anti-Neutrino Detector or \textit{GenericLAND}) GEANT4 simulation for neutrino physics~\cite{glg4sim} and the data analysis framework ROOT~\cite{BRUN1997}.

RAT-PAC models the event-by-event detector response to the signal and background. Events are produced for this study with custom GLG4sim Monte Carlo event generators and particles are propagated in the detector medium with GEANT4. Light emission and PMT response is managed by GLG4sim. RAT-PAC also handles the triggering and data acquisition (DAQ) before the data are output in ROOT format.

The MC model for WbLS is described in more detail in~\cite{Land2021}. The time profile of the scintillation light was based on measurements of WbLS~\cite{Caravaca2020,Onken2020}, and measurements of Gd-WbLS~\cite{Gabriel2022} were used for the light yield and scattering in the model.

For each of the detector configurations simulated, the large-scale, complex structures ({\it e.g.}, I-beams, trusses, PMT support structure and tank) were simplified to approximate position and volume. Each detector was set within a cavern surrounded by a layer of rock with 2~m thickness. For this study, backgrounds from the rock were generated in the inner 10~cm of this layer. This is considered sufficient, in combination with a detector buffer region of 2~m or more and a fiducial volume an additional $\sim 1$~m or more from the PMTs, given that the total neutron flux attenuation is in the region of two orders of magnitude beyond a distance of 3.5~m in rock or water~\cite{Mei2006} and rates of backgrounds due to radioactivity from deeper in the rock are negligible. In the 16~m Gd-WbLS configuration, fewer than 0.5\% of fast neutrons events from the outer 1.9~m of the 2~m rock layer triggered a detector response prior to reconstruction threshold and analysis cuts. In the same simulation, no radioactive decays in the outer layer of rock triggered a detector response.

Event generators were configured for AIT-NEO to produce the initial signal and background particles with spectral and angular distributions as described by a combination of literature and experiment and detailed below.

\subsection{Signals}
\label{sec:signalsims}

The IBD events from the reactor complexes at Hartlepool, Heysham and Torness are simulated using RAT-PAC with spectra from~\cite{Dye2021}. The spectrum of emitted antineutrinos is calculated using monthly-averaged power output and estimated mid-cycle fission fractions. The emitted spectrum is adjusted for the electron antineutrino vacuum survival probability at the distance between the reactor and the detector, assuming the normal neutrino mass ordering. To calculate the theoretically detectable IBD rates from each reactor, the spectral fluxes are multiplied with the IBD cross section given in~\cite{Strumia2003}. 

\begin{figure}[htb]
    \centering
    \includegraphics[width=8.6 cm]{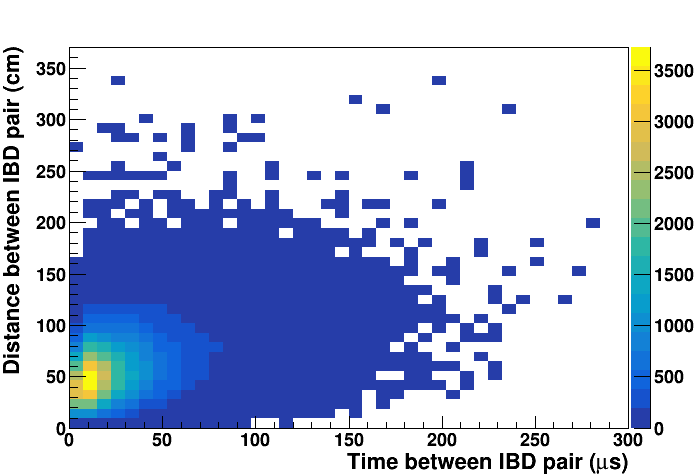}
    \caption{Simulated time and distance between the positron and neutron in an IBD event pair in the 16~m Gd-H$_2$O detector (truth information).}
    \label{fig:ibd}
\end{figure}

Neutrino energies are drawn from the spectral distributions during the simulation to give an appropriate distribution of events for each simulated core or complex. The kinematics of the IBD positron and neutron are defined according to~\cite{Vogel1999}~and~\cite{Strumia2003}. The simulated mean time and distance between pairs in IBD events (Fig.~\ref{fig:ibd}) are $28~\mu s$ and $6$~cm respectively. This coincidence is vital for background rejection.

\begin{figure}[htb]
    \centering
    \begin{subfigure}[b]{8.6 cm}
    \includegraphics[width=\textwidth]{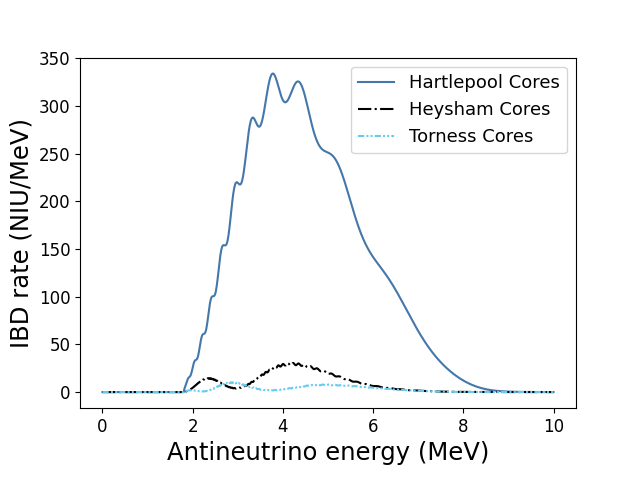}
    \caption{}
    \end{subfigure}
    \hfill
    \begin{subfigure}[b]{8.6 cm}
    \includegraphics[width=\textwidth]{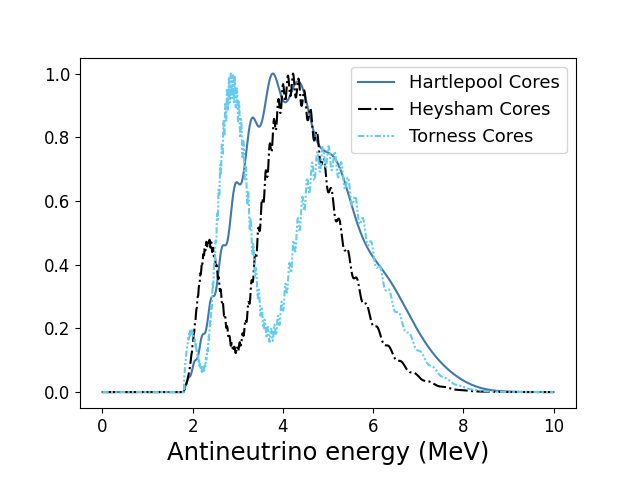}
    \caption{}
    \end{subfigure}
    \caption{IBD spectra at Boulby Mine (a) and normalized spectra (b) at Boulby for the three reactor sites at Hartlepool, Heysham and Torness. The emission spectrum from each reactor is scaled for distance and multiplied with the neutrino oscillation probability and the IBD cross section to produce the number of IBD interactions per MeV per $10^{32}$ free protons per year expected in a detector at Boulby (before detector effects).}
    \label{fig:antinu_spectrum}
\end{figure}

Each Hartlepool core is simulated individually and Heysham is split into two dual-core sites representing the Heysham 1 AGR-1 and Heysham 2 AGR-2 reactors. Torness is simulated as a single dual-core site. Fig.~\ref{fig:antinu_spectrum} shows the total antineutrino IBD spectra at Boulby for these three reactor sites, before detector effects. 

\subsection{Backgrounds}

The overall background rate is significantly reduced by requiring that a pair of prompt and delayed events must occur within the time and distance characteristic of the correlated IBD positron and subsequent Gd neutron-capture signal. However, there remain correlated and uncorrelated background events which can mimic the correlated signal. 

Correlated backgrounds are cosmogenic muon-induced radionuclide and fast neutron backgrounds, as well as IBD backgrounds from other reactors and geoneutrinos. Uncorrelated backgrounds are radioactive decays of isotopes which occur naturally in the detector and its environment. Radioactive isotopes of concern, which are naturally present in the detector components and environs, are detailed in~\cite{Kneale2021}. Illustrative raw event rates for the 16~m baseline detector for each type of background are shown in Table~\ref{tab:background_rates_example}, with associated fractional systematic uncertainties. Uncertainties on the radionuclide backgrounds are also discussed in Section~\ref{sec:radio}. Although many of the $\beta$~decays from ambient radioactivity do not trigger a detector response in simulation, they remain by far the most numerous background events before data reduction.

\begin{table*}[ht!]
    \centering
    \caption{Illustrative raw event rates for each type of background in the 16~m detector and systematic uncertainties on backgrounds. Ambient radioactivity is calculated using rates taken from a combination of data and manufacturer's specifications~\cite{Marti2020,HASELSCHWARDT2019,Hamamatsu2020,Zhang2016b,ARAUJO2012} }
    \begin{ruledtabular}
    \begin{tabular}{lcc}
          Type of background                            &  Rate (Hz)  & Systematic uncertainty\\
        \hline    
         Cosmogenic radionuclide $\beta$-n decays            & $\num{2.0e-5}$  & $<$0.1\%\\
         Cosmogenic fast neutrons                           &  $\num{1.9e-2}$  & 27\%   \\
         Reactor IBD (beyond Hartlepool)                    & $\num{2.2e-06}$  & 6\%   \\
         Geoneutrino IBD                                    &  $\num{3.6e-7}$  & 25\%      \\
         Ambient radioactive isotope $\beta$ decays         &  $\num{3.4e5}$   & Negligible  
    \end{tabular}
    \end{ruledtabular}
    \label{tab:background_rates_example}
\end{table*}

Spontaneous fission of $^{238}$U and $^{232}$Th in the detector medium creates a correlated background of a prompt gamma ray flash and delayed events due to the capture of multiple emitted neutrons. By rejecting events where there are more than two interactions, the event rate for this background can be reduced to $\mathcal{O}(1)$ per year~\cite{Akindele2022}. Backgrounds due to spontaneous fission have therefore been neglected for the purposes of this study.

Cobraa (Section~\ref{sec:Cobraa}) was used to manage the full-detector simulations in RAT-PAC for each of the configurations. For the development and optimization of the analysis for each detector configuration, the signal and each type of correlated background were simulated individually at the relevant expected rates in the detector. Events due to natural radioactivity are combined together into a single simulation where the decays are interleaved at their respective rates as defined by relevant literature~\cite{Marti2020,HASELSCHWARDT2019,Zhang2016b,ARAUJO2012} and manufacturer specifications~\cite{Hamamatsu2020}. 

The backgrounds considered are discussed in more detail in the remainder of this section.

\subsubsection{Radionuclide backgrounds}
\label{sec:radio}

\begin{figure}[htb]
    \centering
    \includegraphics[width=8.6 cm]{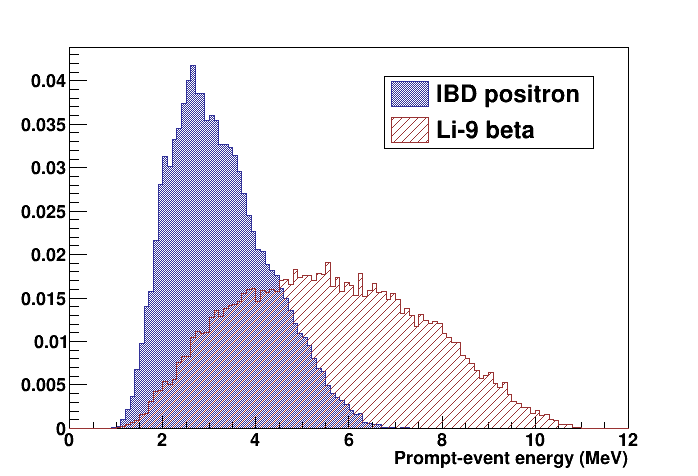}
    \caption[Energy spectra of the IBD positron and $\rm^9Li$ $\beta$]{Normalized energy spectra of the prompt events for IBD (positron) and $\rm^9Li~\beta$-neutron decay ($\beta$ particle). The mean prompt energy in the $\rm^9Li$ decay is higher than that of the positron.}
    \label{fig:max_ep_cut_li9}
\end{figure}

Cosmogenic muons interacting in the detector medium can produce unstable $\beta$-neutron emitters with half-lives of a few seconds in the detector volume. Through-going muons in the detector produce abundant daughter particles that, through successive spallation processes, generate electromagnetic and hadronic showers. Hadronic showers are the principal mechanism by which unstable but relatively long-lived $\beta$-neutron emitting radioisotopes are generated. The decay of these long-lived radionuclides, such as $\rm ^{9}Li$, produces a prompt signal from the $\beta$ particle and a delayed signal from the neutron. These form correlated signals with time intervals and energies similar to the antineutrino signal, although the endpoint energy of the $\beta$ can be higher than the expected energy of the IBD positron - particularly in the case of $\rm ^{9}Li$ (Fig.~\ref{fig:max_ep_cut_li9}).

The radionuclide backgrounds in a water-based Cherenkov detector are due to spallation on oxygen isotopes in water. Table~\ref{tab:radionuclides} shows the spallation-induced radionuclides which are potential backgrounds for reactor antineutrino detection, with yields calculated for the Super-Kamkiokande (Super-K) detector in Japan and consistent with the findings from the WATCHBOY detector~\cite{Dazeley2016}. The radionuclides of particular concern are $^{17}$N and $^{9}$Li due to their relatively high yields combined with $\beta$-n branching ratios as shown.

\begin{table*}[htb]
    \centering
    \caption[Radionuclide production in water]{$\beta$-neutron backgrounds. Isotope yields in water calculated with FLUKA~\cite{Li2014} for Super-Kamiokande. Principal $\beta$ endpoint energies with approximate branching ratios taking into account $\beta$-n branching ratio taken from~\cite{TOI,TUNL,JOLLET2020}.}
    \begin{ruledtabular}
    \begin{tabular}{ccccc}
     Isotope     &  Half life &  Total yield &  Production process & \ $\boldsymbol{\beta}$ endpoints\\
                &  (s)          & {(${10^{-7} {\mu}^{-1}\rm g^{-1}~cm^2}$)}  & &  (MeV)\\
    \hline
    $^{17}$N            &   4.173    &    0.59   & $^{18}$O(n,n+p)       & 3.3 (50\%), 4.1 (38\%)\\
    $^{16}$C            &   0.747    &    0.02    & $^{18}O$($\pi^-$,n+p)         & 4.7(84\%), 3.7(16\%)\\
    $^{11}$Li (85\% br) &   0.0085   &    0.01    & $^{16}O$($\pi^+,5p+\pi^++\pi^0$)   & 16.6(22\%)$^*$, 12.5(16\%)\footnote{Total branching ratio.} \\
    $^9$Li (50.8\% br)    &   0.178    &    1.9   & $^{16}O$($\pi^-$,$\alpha$+2p+n) & 11.2(29\%), 10.8(12\%)\\
    $^8$He (16\% br)    &   0.119    &    0.23  & $^{16}O$($\pi^-$,3H+4p+n)       & 5.3(8\%), 7.5(8\%)\\
    \end{tabular}
    \end{ruledtabular}
    \label{tab:radionuclides}
\end{table*}

The total muon flux in Boulby mine is $\mathit{ \Phi}_\mu = (4.09 \pm 0.15) \times 10^{-8} \rm~cm^{-2} ~s^{-1}$~\cite{Robinson2003}. The $\beta$-neutron decay rates of the radionuclides of concern are calculated according to Equation~\ref{eq:rn_rates}:

\begin{equation}
\begin{aligned}
 R_{iso} (\mathrm{s^{-1}}) = R_\mu &\times L_\mu (\mathrm{cm}) \times  Y_{iso} (\mu^{-1}~\mathrm{g^{-1}~cm^2}) \\
  & \times br \times \rho (\mathrm{g ~cm}^{-3}) \times \bigg(\frac{E_{\mu,Boulby}}{E_{\mu,Super-Kamiokande}}\bigg)^{\alpha},
\label{eq:rn_rates}
\end{aligned}
\end{equation}
where $\mathit{ R}_\mu = \mathit{\Phi}_\mu \times$ tank surface area, $\mathit{L}_\mu$ is the muon path length, $\mathit{ Y_{iso}}$ is the isotope yield, $\mathit{br}$ is the branching ratio, and $\rho = \rm 1~g~cm^{-3}$ for water. The muon path length is taken to be the vertical height of the detector in the calculation of the yields, which makes the simplifying assumption that the muons are all downward going. There is a depth-related correction for the average muon energy of $\mathit{E}_\mu^{\alpha}$ where $\alpha = 0.73 \pm 0.10$~\cite{Mei2006} since, on average, higher-energy muons will survive to greater depths.

Although measurements of the isotope yields have been made~\cite{Zhang2016}, there was a large disparity between theory and experiment and so the theoretical values were adopted for this study as a cautious approach, since they give higher rates. The principal radionuclide backgrounds $^{17}$N and $^{9}$Li were included.

The uncertainty on the rates is
\begin{equation}
    \frac{\sigma_{R_{iso}}}{R_{iso}} = \sqrt{\frac{\sigma_{\Phi_\mu}^2}{\Phi_\mu} + \frac{\sigma_{E_{\mu}^\alpha}^2}{E_{\mu}^\alpha}},
\end{equation}
where $\sigma_{\mathit{\Phi}_\mu} = 0.15 \times 10^{-8}$ and $\sigma_{E_\mu^\alpha}$ are the uncertainties on the muon flux and energy dependence respectively and
\begin{equation}
\sigma_{E_\mu^\alpha} = \sqrt{(E_\mu^\alpha\ln(E_\mu) \sigma_\alpha)^2 + (\alpha E_\mu \sigma_{E_\mu})^2},
\end{equation}
where $\sigma_{E_\mu}$ is due to uncertainties in the atmospheric muon energy spectrum and muon energy loss in the rock but is not highly correlated with the rock composition or precise depth~\cite{Robinson2003} and so is assumed to be negligible for this work. It should be noted that although the calculated systematic uncertainty on the rate from the theoretical yield is very low (Table~\ref{tab:background_rates_example}), adopting the theoretical yield remains a cautious approach.

\subsubsection{Fast Neutrons}

Cosmogenic muons interacting outside the detector ({\it e.g.}, in the rock) induce fast neutrons by muon spallation and also neutron evaporation along a muon track. It is common for these neutrons to be produced in multiplicity, which can generate neutron pairs in the detector volume. Fast neutron pairs in the detector volume thermalize and capture, producing a detector response which mimics the reactor antineutrino signal. 

\begin{figure}[htb]
    \centering
    \begin{subfigure}[b]{8.6cm}
    \includegraphics[width=\textwidth,height=60mm]{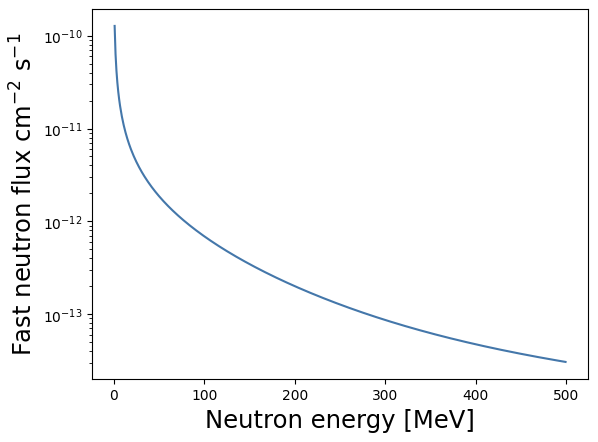}
    \caption{}
    \end{subfigure}
    \hfill
    \begin{subfigure}[b]{8.6cm}
    \includegraphics[width=\textwidth,height=60mm]{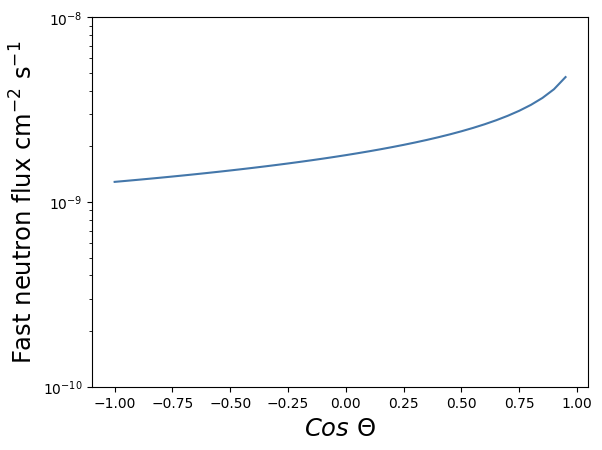}
    \caption{}
    \end{subfigure}
    \caption[Fast neutron energy and angular distributions at Boulby]{Fast neutron energy spectrum (a) at Boulby in the range from 0 to 500 MeV~\cite{Wang2001} and neutron angular distribution (b) with respect to the muon direction for an average muon energy of 264 GeV~\cite{Mei2006}.}
    \label{fig:fn}
\end{figure}

The fast neutron spectrum is simulated as defined in~\cite{Wang2001}:

\begin{equation}
    \frac{dN}{dE_n} = A \big(\frac{e^{-0.7E_n}}{E_n}+B(E_\mu)e^{-2E_n} \big),
\end{equation}
where \textit{A} is a normalization factor and $\mathit{B}(E_\mu) = 0.52 - 0.58e^{-0.0099E_\mu}$ is a function of the muon energy \textit{E}$_\mu$. An alternative definition is given in~\cite{Mei2006}. Fast neutrons from the rock below 10~MeV are not considered to be of concern in this case as they are insufficiently penetrating. However, a significant number of fast neutrons have energies greater than 10~MeV (Fig.~\ref{fig:fn}), with an average neutron energy of 88 MeV and an average muon energy at Boulby of 264~GeV~\cite{Mei2006}. The fast neutron flux in the Boulby Mine at the cavern wall interface, calculated through a combination of simulation and experimental data for fast neutrons above 10 MeV~\cite{Mei2006}, is
\begin{equation*}
    \rm \Phi_n(>10~MeV) = 1.11 \times 10^{-9} ~cm^{-2} s^{-1} \times cavern ~surface ~area.
\end{equation*}
Note that there is a very high uncertainty of 27\% on this flux due to uncertainties on the muon flux and neutron production rates.

Cosmogenic muons are strongly biased in the downward direction, particularly at the high energies required to penetrate underground to the depth of the cavern at Boulby. The muon angular distribution with respect to the Earth's surface at the detector location is simulated according to the distribution given in~\cite{Tang2006}. Attenuation of muons is simulated to vary as a function of this muon angle with the distance traveled through the Earth and is also described in~\cite{Tang2006}. 

Fast neutrons from muon spallation are peaked in the direction parallel to the muon direction. Secondary neutrons from neutron evaporation (nuclear de-excitation with neutron emission) along the path of the muon are emitted isotropically~\cite{Wang2001}. The resulting neutron angular distribution with respect to the muon direction (Fig.~\ref{fig:fn}) is peaked in the direction of the muon, with a flattening effect due to the secondary neutrons. The highest-energy neutrons tend to be emitted in the same direction as the muon but lower-energy neutrons ($<$100~MeV) are more isotropically distributed. The highest-energy, downward-going neutrons are most easily rejected, since they will be accompanied by a very high-energy muon track in the detector. The more isotropic, lower-energy and secondary neutrons are of more concern.

\begin{figure}[htb]
    \centering
    \includegraphics[width=8.6 cm]{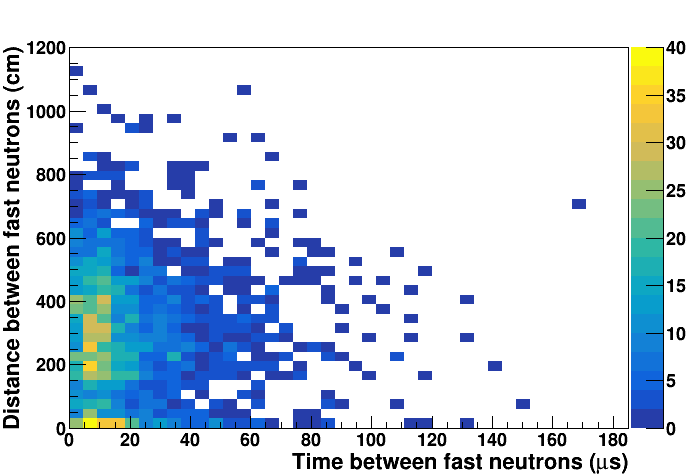}
    \caption{Simulated time and distance between consecutive fast neutrons in Gd-H$_2$O in the 16~m cylinder (truth information). The mean time betwen fast neutrons ($\rm 20~\mu s$) is similar to that of IBD pairs and many occur within the spatial coincidence range of the IBD pair.}
    \label{fig:fn_dTdR}
\end{figure}

The simulated mean time and distance between consecutive fast neutron events are $\sim20~\mu \rm s$ and $\sim300 \rm~cm$ respectively in Gd-H$_2$O  (Fig.~\ref{fig:fn_dTdR}). The time coincidence of fast neutron pairs is similar to that of IBD event pairs, and many of the fast neutron events occur within the spatial range of the IBD pairs.

The neutron yield per muon is dependent on the muon residual energy and thus also on the depth, since higher-energy muons are more likely to penetrate deeper underground. It is additionally dependent on the material in which the muon is interacting. A hundred or more coincident neutrons may be produced by a muon in the rock at Boulby, although the majority of muons result in less than twenty coincident fast neutrons and the average expected multiplicity at Boulby is 6.03 per muon, given an assumed rock density of $\rm2.7 g~cm^{-3}$~\cite{Mei2006}. Measurements with the WATCHBOY and Multiplicity and Recoil Spectrometer (MARS) detectors found the observed rate of two correlated fast neutrons to be consistent with the rate predicted by simulation~\cite{Sutanto2020}.

\subsubsection{IBD Backgrounds}

Antineutrinos from reactor cores other than the signal core(s) produce a true IBD background which will not be distinguishable from the signal under most circumstances. The reactor background at Boulby depends on the particular signal being observed. Geological electron antineutrinos, commonly referred to as geoneutrinos, emanate principally from $\beta$~decays of the $^{238}$U and $^{232}$Th decay chains and $^{40}$K in the Earth's crust and mantle.  Where they have an energy over the IBD threshold, they can interact in a water-based detector to form a subdominant IBD background to reactor antineutrinos.

\begin{figure}[htb]
    \centering
    \begin{subfigure}[b]{8.6 cm}
    \includegraphics[width=\textwidth]{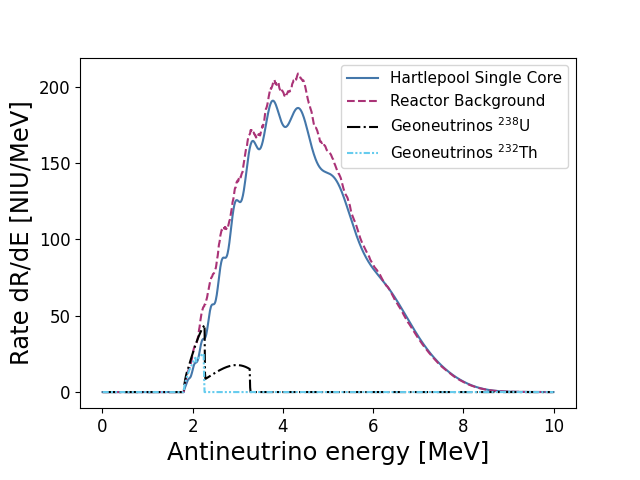}%
    \caption{}
    \end{subfigure}
    \hfill
    \begin{subfigure}[b]{8.6 cm}
    \includegraphics[width=\textwidth]{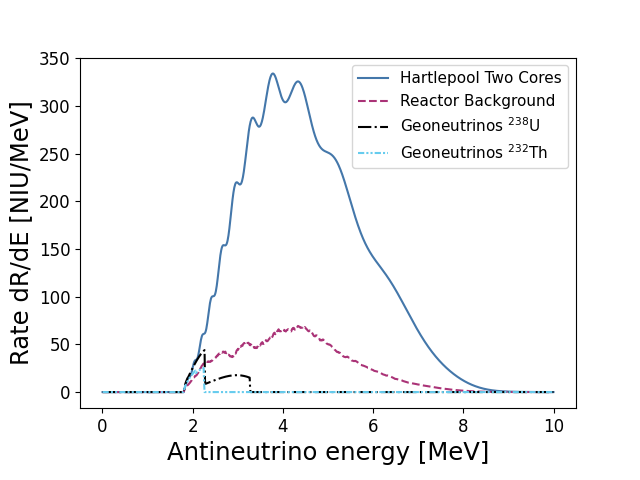}%
    \caption{}
    \end{subfigure}
    \caption[Background spectra at Boulby]{Background IBD spectra at Boulby for a single Hartlepool core (a) and both Hartlepool cores (b) (before detector effects). Reactor backgrounds include IBD interactions from antineutrinos emitted by all nearby and world reactors excluding the signal reactor(s). Geoneutrinos are a sub-dominant IBD background for all signals investigated. }
    \label{fig:antinu_backgrounds}
\end{figure}

The reactor background at Boulby will depend on the particular signal being observed. Fig.~\ref{fig:antinu_backgrounds} shows the reactor background at Boulby for a single-core (a) and twin-core (b) signal at Hartlepool. The background to the single-core signal includes the second core. The reactor backgrounds exceed the signal in the single-core case. The spectra and rates are calculated in a similar way to the signal and summed over all other reactors.

Geoneutrino spectra (Fig.~\ref{fig:antinu_backgrounds}) are also taken from~\cite{Dye2021}. The geoneutrino spectrum for each isotope is estimated by summing the spectra from each $\beta$ decay in the chain, weighted by its branching ratio. The geoneutrino IBD rates for the detectors are calculated by multiplying the geoneutrino spectral flux by the IBD cross section and neutrino vacuum survival probability (assuming normal ordering) over the spectrum. Adjustments are made for variations in the density of the mantle and natural abundances of each isotope.

\subsubsection{Uncorrelated backgrounds}

An accidental coincidence of two physically independent interactions due to natural levels of radioactivity in the detector components and environment can mimic the time-correlated IBD signal.

The rate of accidental coincidences depends on the composition of the detector (PMTs, detector medium, steel) and cavern in which the detector is situated (concrete, surrounding rock and air). It also depends on the energy threshold, vertex position, vertex resolution, and the distance of the fiducial volume from the detector and cavern materials.  Isotopes which decay with a non-negligible half-life and energy around or above the Cherenkov threshold are considered a background. Radioactive isotopes naturally present in the detector components and environs and the nominal radioactivity currently adopted for AIT-NEO are given in~\cite{Kneale2021}.

\begin{figure}[hb]
    \includegraphics[width=86mm]{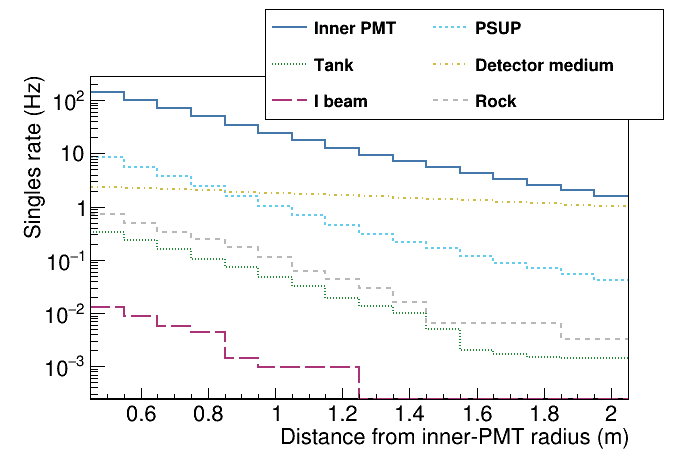}
    \caption[Radioactive backgrounds by component in the detector volume]{Single events due to natural radioactivity of components in and around the detector which can result in accidental coincidences and mimic the signal. Results for the 16~m tank with a Gd-WbLS fill are shown but relative rates in Gd-H$_2$O and the larger tank are similar.}
    \label{fig:singles}
 \end{figure}

The principal contribution to backgrounds from accidental coincidences comes from $^{238}$U and $^{232}$Th in the PMT glass (Fig.~\ref{fig:singles}) and using PMTs made with low-radioactivity glass can help to mitigate this background. Low-radioactivity glass has been adopted for this study. In addition, many of these events can be rejected by applying a fiducial cut, which rejects all events within a given distance from the PMTs.

Supported radon in the medium, due to the emanation of radon from the decay of $^{238}$U in the PMTs and absorption from the mine air, plus other radioactivity in the medium, tends to be the next most significant background once a fiducial cut has been applied, since the fiducial cut eliminates many of the contributions from the steel of the PMT support structure. The radon levels in Boulby Mine are comparatively low at 3~Bq~m$^{-3}$. Nevertheless, the rate of radon diffusion from glass into water is not well known and these backgrounds occur throughout the detector volume, making them difficult to reject.

Only $\beta$ decays are considered. Alpha decay products are not relativistic and are therefore not visible in a Gd-$\rm H_2 O$ Cherenkov detector. Similarly, they are not expected to contribute to backgrounds in Gd-WbLS due to their rapid attenuation. Decays with a branching fraction below 0.1\% and/or the endpoint energy below 0.5 MeV, as well as decays with a triggered singles rate of $<10^{-3}$ Hz are neglected, since these were found to rarely contribute to accidental coincidences.

\section{Sensitivity Evaluation}
\label{sec:sensitivity}

Two parallel analysis paths - \textit{Cobraa} and \textit{LEARN} - have been developed to test the sensitivity of an antineutrino detector to remote reactor signals. The separate analysis chains provide a validation mechanism by means of a cross check of independently obtained results. 

The two analysis pathways begin with shared Monte Carlo simulations described in Section~\ref{sec:sims}. After the Monte Carlo is generated, the two paths diverge, with the exception of an analytical post-muon veto described in Section~\ref{sec:muon-veto}. The two analyses are described in Section~\ref{sec:Cobraa} and Section~\ref{sec:Learn}. 

\subsection{Sensitivity Metric}
\label{subsec:sensitivity-metric}

The sensitivity of the experiment is expressed as the expected required observation time to detect an unknown reactor where backgrounds are known. This is referred to as the experiment \textit{dwell time}. The dwell time is the sensitivity metric used for comparison of the detector configurations. The dwell times are quoted under the assumption that the reactors have the same average power output as quoted by the International Atomic Energy Agency (IAEA) for the year 2020, and there are no abnormal shutdowns. This takes into account regular maintenance and refueling schedules.

Two levels of detection are considered - \textit{anomaly measurement} and \textit{reactor measurement}.

\subsubsection{Anomaly Measurement}

The most simple detection of an unknown reactor is taken to be rejection of the null hypothesis that the rates are consistent with expected background rates. This is termed an anomaly measurement. Rejection of the null hypothesis to $3\sigma$ (or equivalent) significance is considered to be sufficient in the application to non-proliferation for this type of scenario. Two sensitivity metrics have been derived to test this background-only hypothesis, which are selected according to the strength of the signal relative to the background, as explained below.

The sensitivity metric for an anomaly measurement of one or more reactor core(s) in the presence of known backgrounds is most simply derived from the Gaussian significance:
\begin{equation}
    N_\sigma = \frac{s~t}{\sqrt{\sum\limits_{i=0}^{n} b_it + \sum\limits_{i=0}^n (\sigma_{b_i}t)^2}},
    \label{equ:gaussian_anomaly}
\end{equation} 
where $\mathit{N_\sigma}$ is the number of Gaussian standard deviations from the expected background counts, $s$~$t$ is the expected total signal counts at rate \textit{s} in time \textit{t}, $\mathit{b_i}$ is the expected rate of the $\mathit{i^{th}}$ background, $\mathit{\sum_{i=0}^n b_it}$ is the expected total background counts summed over \textit{n} backgrounds and $\mathit{\sum_{i=0}^n\sigma_{b_i}t}$ is the total systematic uncertainty on all backgrounds. This gives an equation for the dwell time as follows:
\begin{equation}
    t_{dwell} = \frac{N_\sigma^2\sum\limits_{i=0}^n b_i}{s^2 - N_\sigma^2\sum\limits_{i=0}^n\sigma_{b_i,sys}^2},
    \label{equ:gaussian_anomaly_dwell}
\end{equation}
which is calculated for $\mathit{N}_\sigma = 3$.

This approximation is valid where $\mathit{s \ll b}$ and where the number of counts is greater than 20 to 30. Where these criteria are not met, the Gaussian approximation is replaced with a Poisson distribution and the total systematic uncertainty on the background is incorporated as a Gaussian-distributed nuisance parameter. The resulting expression for the Poisson significance \textit{Z}, derived in~\cite{Vianello2018}, is more complex than Equation~\ref{equ:gaussian_anomaly}:
\begin{equation}
   Z =  \Bigg[-2(s\cdot t+\sum\limits_{i=0}^n b_it) \ln \frac{\hat{B}_0}{s\cdot t+\sum\limits_{i=0}^n b_it} - \frac{(\sum\limits_{i=0}^n b_it- \hat{B}_0)^2}{2 \sum\limits_{i=0}^n(\sigma_{b_i} t)^2} - \hat{B}_0 + s\cdot t + \sum\limits_{i=0}^n b_it \Bigg]^\frac{1}{2},
   \label{equ:sig_poiss_gauss_main}
\end{equation}
where the background rate $\mathit{ \hat{B}_0}$ for $\mathit{s}$ $\mathit t=0$ is:
\begin{equation}
\begin{aligned}
\mathit{\hat{B}_0} = & \frac{1}{2}\Bigg(\sum\limits_{i=0}^n b_it-\sum\limits_{i=0}^n (\sigma_{b_i}t)^2 + \\ &\sqrt{\sum\limits_{i=0}^n (b_it)^2-2\sum\limits_{i=0}^n b_it \sum\limits_{i=0}^n(\sigma_{b_i}t)^2+4(s \cdot t+\sum\limits_{i=0}^n b_it) \sum\limits_{i=0}^n(\sigma_{b_i}t)^2+ \sum\limits_{i=0}^n (\sigma_{b_i}t)^4}~\Bigg).
\end{aligned}
\end{equation}

Equation~\ref{equ:sig_poiss_gauss_main} for the significance can be expressed as a function of a free parameter \textit{t} and in this way the value for $\mathit{t_{dwell}}$ can be extracted for the required significance \textit{Z=3}, where \textit{Z} is the Poisson equivalent of the Gaussian $\mathit{N_\sigma}$.

\subsubsection{Reactor Measurement}

A more challenging detection of an unknown reactor is taken to be measurement of rates consistent with signal plus background rates. This is termed a reactor measurement. Confirmation of the signal-plus-background hypothesis to $3\sigma$ significance is again considered to be sufficient for the reactor measurement scenario.

The sensitivity metric for a measurement of one or more reactor core(s) in the presence of known backgrounds is most simply derived from the Gaussian significance:
\begin{equation}
    N_\sigma = \frac{s~t}{\sqrt{s~t + (\sigma_s t)^2 + \Sigma b_it + \Sigma (\sigma_{b_i}t)^2}}
    \label{equ:gaussian_measurement}
\end{equation} 
where $\sigma_s$ is the systematic uncertainty on the signal rates, taken to be equal to the systematic uncertainty on the reactor backgrounds as summarized in Table~\ref{tab:background_rates_example}. The resulting dwell time calculation is
\begin{equation}
    t_{dwell} = \frac{N_\sigma^2(s + \Sigma b_i)}{s^2 - N_\sigma^2(\sigma_s^2 + \Sigma\sigma_{b_i,sys}^2)},
    \label{equ:gaussian_measurement_dwell}
\end{equation}
with $\mathit{N}_\sigma = 3$. 

\subsection{Cobraa pathway}
\label{sec:Cobraa}

The Cobraa pathway uses a Coincidence Reconstruction (CoRe) for vertex reconstruction, followed by a multivariate analysis to optimize detector sensitivity.

CoRe was developed specifically to capitalize on the coincidence of events in a gadolinium-doped detector medium, by reconstructing pairs of events together. It combines the detected light from both the positron and neutron to improve the vertex reconstruction. This exploits the fact that the positron and neutron vertices are sufficiently close together (a mean true distance of 6~cm), within the expected vertex resolution, as to be considered the same.

Reconstruction in the standard AIT-NEO tools is achieved using BONSAI (Branch Optimization Navigating Successive Annealing Iterations), which is a maximum likelihood fitter to the timing and spatial pattern of the PMT hits. BONSAI has been used for many years to perform low-energy reconstruction up to 100 MeV for Super-Kamiokande~\cite{Smy2007} and has since been optimized and implemented for AIT-NEO. The CoRe combined vertex reconstruction is an extension of BONSAI to include a combined fit, which maximizes a combined likelihood for the prompt and delayed event to output a single vertex for a given pair of events. CoRe is described in detail in~\cite{Kneale2021,Kneale2022}.

The combined fit improved the vertex resolution for IBD positrons and neutrons compared to the BONSAI single-event fit for all detector configurations studied. Improvements at lower energies in particular can improve overall sensitivity and help to lower the reconstruction threshold. An additional benefit of the combined fit is to provide an effective neutron-tagging method, which helps with the rejection of accidental coincidences via a cut on the quality of the fit (Section~\ref{subsubsec:accidentals}).

The new simulation and analysis tool Cobraa (Coincident-Background Reactor Antineutrino Analysis)~\cite{Cobraa2021} was developed to perform a full evaluation of coincident signal and background events for reactor antineutrino detection. Cobraa was originally developed within the framework of the WATCHMAKERS simulation and analysis tool developed for AIT-NEO~\cite{WATCHMAKERS}. Using combined simulations, Cobraa performs a full evaluation of accidental coincidences of uncorrelated backgrounds and of true coincidences of the correlated signal and background events.

The Cobraa-CoRe toolchain takes the user from simulation in RAT-PAC of coincident signal along with correlated and uncorrelated backgrounds as detailed in Section~\ref{sec:sims}, through the CoRe combined vertex fitting, to parallel optimization of the dwell time via analysis cuts in six dimensions.

\subsubsection{Optimization of Dwell Time}
\label{cobraa:optimisation}

The Cobraa-CoRe analysis optimizes the dwell time defined in Section~\ref{subsec:sensitivity-metric} via data-reduction cuts in six dimensions:
\begin{enumerate}
    \item Fiducial cut - rejects all events below a minimum distance from the inner PMT radius.
    \item Prompt-event energy threshold cut E$_{p,min}$ - rejects pairs with a lower prompt-event energy.
    \item Delayed-event energy threshold cut E$_{d,min}$ - rejects pairs with a lower delayed-event energy.
    \item Cut on maximum time between events dT$_{max}$ - rejects pairs where the time between the two events $\mathrm{dT>dT}_{max}$.
    \item Minimum \textit{timing goodness} g$_{min}$ - rejects poorly reconstructed events.
    \item Maximum prompt-event energy E$_{p,max}$ - rejects pairs with a higher prompt-event energy than expected from an IBD positron.
\end{enumerate}

The fiducial cut is applied as a minimum distance from the PMTs and creates a virtual inner target volume (fiducial volume) within the inner detector. For example, a 1~m fiducial cut would create a fiducial volume with a 4.7~m radius in the 16~m detector and an 8~m radius in the 22~m detector. This helps to reduce backgrounds due to fast neutrons and radioactivity in the PMTs and steel structures.

The amount of light produced by an event is dependent on the energy of the event. As a result, an energy analogue - the number of PMT hits due to unscattered light from a single event - is used for the cuts on energy. Only PMT hits in a narrow time window are selected in order to filter out hits from dark noise and scattered light. To maximize the number of hits from unscattered light and minimize the dark hits and scattered light, a time window of 9~ns is used for Gd-H$_2$O - 3~ns before the peak of the light from a single event and 6~ns after. A time window of 100~ns (10~ns before and 90~ns after) is used for Gd-WbLS. The \textit{n9} and \textit{n100} energy analogues are therefore the number of PMT hits with a residual time within a 9~ns or 100~ns time window around the peak of the light. For the 22~m detector with a Gd-H$_2$O fill, it was found that the n9 estimator was not optimal. For this reason n100 has been used for the 22~m Gd-H$_2$O detector. Optimization of the values for the larger detector and the Gd-WbLS fill would need to be performed to take into account expected noise effects in real data. 

The timing goodness is a measure of the fit quality developed for BONSAI~\cite{Smy2007} and also output by CoRe. It is a measure of the coincidence of the \textit{time residuals} (time of hit - time of flight from reconstructed vertex - reconstructed time of emission) from the reconstructed vertex. It is effective as a measure of fit quality in that the coincidence will be poor if the reconstruction is poor. A pair of uncorrelated events which are far apart in distance are likely to have a poor measure of quality for a fit which outputs a single vertex for both events, such as CoRe. For this reason, a cut on the fit quality can be viewed, in part, as an analogue for a cut on the distance between events.

\begin{figure}[htb]
    \includegraphics[width=86mm]{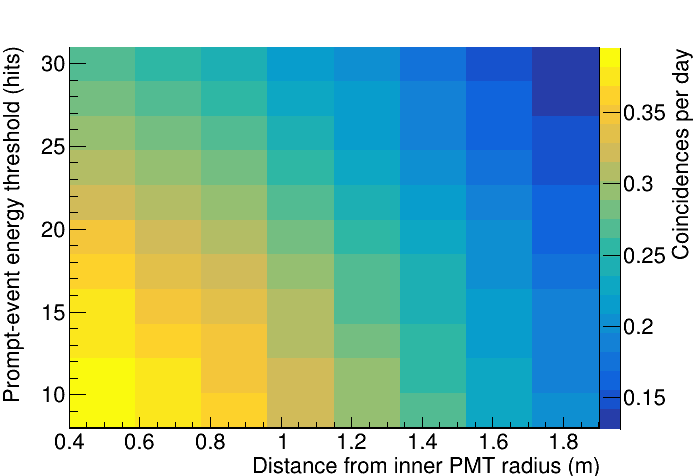}
 \caption[Signal coincidences per day from coincidence evaluation in Cobraa analysis]{Two-dimensional profile output by Cobraa showing the coincident-signal rates per day in Gd-WbLS in the 16~m detector. The fiducial cut and prompt energy threshold are allowed to vary while the other cuts are held constant. The highest signal (lightest, yellow) is at the lowest energy and with the smallest fiducial cut (largest fiducial volume). The signal rate decreases as each of these cuts increase. The most aggressive cuts result in the lowest signal (darkest, blue).}
    \label{fig:signal_coincidence}
\end{figure}

The first stage of the analysis is to evaluate the rate of coincidences for each combination of cuts. An additional multiplicity cut is applied to reduce the fast neutron background (Section~\ref{coincidence:cosmics}). Where the number of coincidences is evaluated to zero, the 95\% Poisson~\footnote{Strictly, the binomial upper confidence level is more appropriate here but the Poisson upper confidence limit gives a more conservative estimate of the uncertainty.} upper confidence limit is used. The coincidence rates are output in two-dimensional slices as a function of prompt-event energy threshold and fiducial cut for each combination of the remaining cuts. Fig.~\ref{fig:signal_coincidence} shows one such slice for the IBD signal, where the number of IBD coincidences decreases as the cuts increase.

The second stage of the analysis performs the final sensitivity optimization. It reads in the rates from the previous step, adjusts the radionuclide rates for an analytical post-muon veto (Section~\ref{sec:muon-veto}) and evaluates the signal significance and associated uncertainties at each combination of cuts to find the optimal signal significance and associated optimal cuts, rates and uncertainties.

Optimizing on the dwell time has the effect of optimizing automatically on either the signal over background significance or the background and uncertainty on the background, where statistics or systematic uncertainties are the limiting factor respectively.

\subsubsection{Rejection of Accidental Coincidences}
\label{subsubsec:accidentals}

The first five of the cuts over which Cobraa optimizes, and which are described above, are the most effective for the rejection of accidental coincidences. Fig.~\ref{fig:singles} in Section~\ref{sec:sims} shows that a fiducial cut to remove events close to the PMTs is highly effective in reducing many of the uncorrelated radioactive-decay events.

The prompt and delayed energy thresholds can remove more than 99\% of the accidental coincidences which pass the DAQ trigger and reconstruction thresholds. Separate energy thresholds for the prompt and delayed events increases the rejection of accidental coincidences in Gd-H$_2$O by allowing the energy threshold of the delayed, neutron capture-like event to be higher, while preserving as much of the positron signal as possible. This effect increases with increasing tank size. The effect also increases as the signal to background ratio decreases in order to maximize background rejection. In Gd-WbLS, the prompt energy threshold tends to decrease as the detector size increases and increase as the signal to background ratio decreases, while the delayed-event energy threshold decreases as the signal to background ratio decreases.

\begin{figure}[htb]
    \includegraphics[width=86mm]{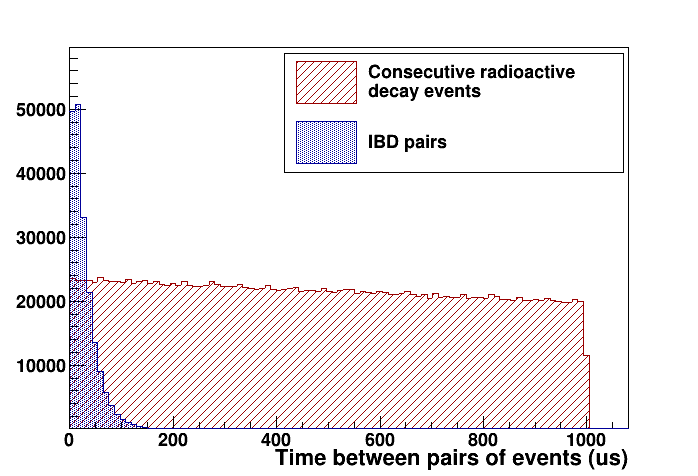}
 \caption{Typical time between consecutive uncorrelated events due to radioactivity compared to time between IBD positron-neutron pairs. A cut-off at 1000~$\rm \mu$s has been applied to the time between consecutive uncorrelated events but in fact the mean time between events is 6000~$\rm \mu$s and the distribution is an exponential which extends to tens of 1000s of $\mu$s.}
    \label{fig:singles_dT_dR}
\end{figure}

Many of the remaining events are removed by the cuts on the time between events and the fit quality. Fig.~\ref{fig:singles_dT_dR} shows the typical time and distance between consecutive uncorrelated events. Rejecting events separated by more than $\rm 100~\mu s$ in Gd-H$_2$O removes more than 50\% of the uncorrelated events which pass the trigger threshold. The optimization is carried out over a range of time cuts around a value close to the expected maximum time between the positron and neutron events in an IBD event pair. The optimal time between events is independent of detector size but increases with the signal to background ratio. In Gd-H$_2$O, the optimal time is $\rm \sim120~\mu s$ for the near reactors and $\rm \sim200 ~\mu s$ for the more distant reactor signals. The optimal time between events in Gd-WbLS is over $\rm 200 ~\mu s$.

\begin{figure}[htb]
  \includegraphics[width=86mm]{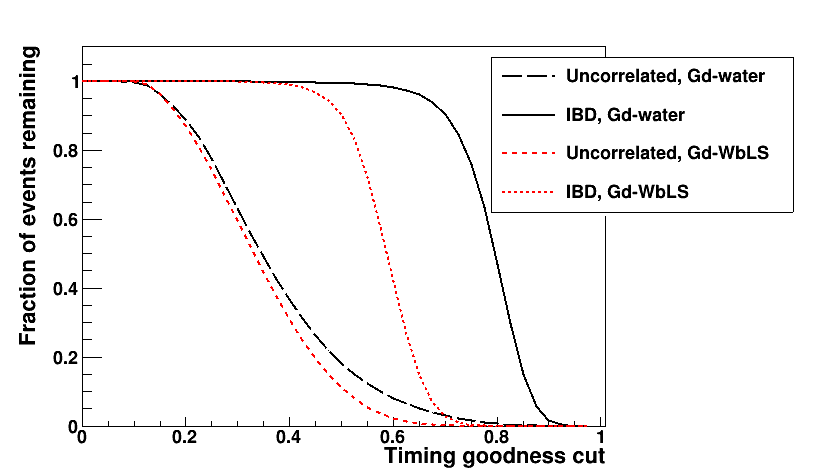}
\caption[Effectiveness of fit quality threshold for correlated and uncorrelated events]{Fraction of correlated IBD event pairs for Gd-H$_2$O (black, solid) and Gd-WbLS (red, dotted) and accidental coincidences of uncorrelated radioactive decays for Gd-H$_2$O (black, long-dashed) and Gd-WbLS (red, short-dashed) remaining as a function of a fit quality threshold as measured by the timing goodness.}
    \label{fig:eff_goodness}
 \end{figure}

Optimization of the cut on the minimum timing goodness exploits the false-pair rejection power of CoRe. Accidental coincidences of uncorrelated events tend to result in a poorer vertex reconstruction in CoRe. Consequently, the timing goodness rejects false pairs which pass the time coincidence cut. Optimal timing goodness threshold cuts tend to be 0.4 in Gd-WbLS and 0.6 in Gd-H$_2$O. Fig.~\ref{fig:eff_goodness} shows that cuts at these values can reject more than 50\% of the uncorrelated events while retaining all of the signal events. 

The combination of the CoRe reconstruction and subsequent Cobraa analysis tends to remove all of the accidental coincidences in Gd-H$_2$O. In this case, an upper confidence limit is given. Even in Gd-WbLS, which has a lower energy threshold and `sees' more of the uncorrelated events, accidental coincidences are reduced to a subdominant background.

\subsubsection{Rejection of muon-induced backgrounds}
\label{coincidence:cosmics}

\begin{figure}
    \centering
    \includegraphics[width=8.6 cm]{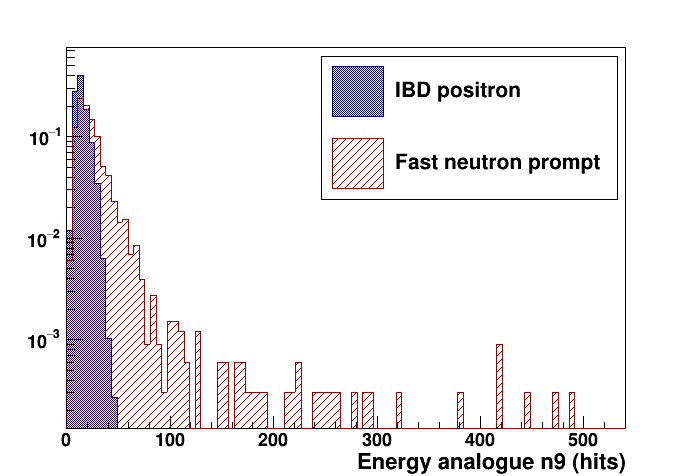}
    \caption{Energy analogue for the prompt events in IBD and fast neutron pairs which occur within $200~\mu s$ of each other.}
    \label{fig:n9_fn_ibd}
\end{figure}

In a fast neutron pair, the first (prompt) event is likely to be, on average, much higher in energy than a positron in an IBD event (Fig.~\ref{fig:n9_fn_ibd}). A maximum prompt-event energy cut ($E_{p,max}$), which is optimized in a range between the peak positron energy and the maximum positron energy, is an effective tool in rejecting fast neutron events where the prompt-event energy is not typical of a positron. 

The optimal $E_{p,max}$ cut tends to be higher in Gd-WbLS. In both media, the cut decreases as the signal to background ratio decreases.

Additional fixed cuts were developed to deal with the fast neutron background:
\begin{itemize}
    \item Re-trigger threshold cut
    \item Multiplicity cut
\end{itemize}

A threshold of 1~$\mu$s between events helps to reject re-triggers. These are particularly problematic at the very high energies of fast-neutron events. 

Fast neutron events can consist of up to (and sometimes in excess of) eight coincident events in the detector. Rejecting multiply-coincident events can therefore help to reject fast neutron events. The Cobraa analysis accepts only pairs of events which have no other coincident event before or after the pair. This can reduce the fast neutron rate in the fiducial volume by greater than 90\%.

\begin{figure}[htb]
  \includegraphics[width=75mm]{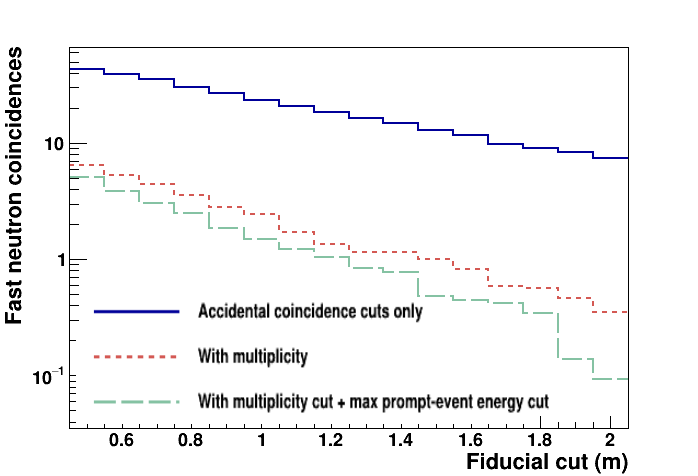}
 \caption[Fast neutron rejection in the Cobraa analysis]{Fast neutron coincidences in Gd-H$_2$O in the 16~m detector as a function of fiducial cut in terms of distance from the PMTs, with the `basic' cuts for rejection of accidental coincidences (blue, solid), additional cut on multiplicity (red, short-dashed), and additional cut on multiplicity plus a fixed $E_{p,max}$ of 40 \textit{n9} hits (green, long-dashed).}
    \label{fig:fast_neutron_rejection}
\end{figure}

The effectiveness of the multiplicity and maximum prompt-event energy cut-off for fast neutrons is evident in Fig.~\ref{fig:fast_neutron_rejection}, which gives an indication of the relative numbers of events remaining, as a function of the fiducial cut expressed as the distance from the inner PMT radius, after the multiplicity and maximum prompt-event energy cuts have been applied.

The $E_{p,max}$ cut is also effective in reducing the radionuclide backgrounds. It is particularly useful in the case of $\rm ^9Li$, where the $\beta$ tends to be higher in energy than the IBD positron, as seen in Section~\ref{sec:radio}. The final stage in the reduction of muon-induced backgrounds is to apply an analytical post-muon veto as described in Section~\ref{sec:muon-veto}, which rejects many of the remaining radionuclide backgrounds.

\subsubsection{Optimized cuts}

Optimised cuts are given in Tables~\ref{tab:cobraa_cuts_water}~and~\ref{tab:cobraa_cuts_wbls}. Increasing the detector size improves sensitivity for a given fill by effectively increasing the size of the fiducial volume, which increases the signal rate. Although some background rates also scale with size, the higher signal rate makes it possible to cut more aggressively on the fiducial volume and energy cuts, resulting in proportionately better rejection of background rates. This results in a net increase in sensitivity and a decrease in dwell time with increasing detector size. 

Upgrading the detector fill from Gd-H$_2$O to Gd-WbLS improves sensitivity to the signal thanks to the increase in light, particularly at the lower end of the reactor spectrum. This also increases sensitivity to backgrounds - again, particularly at lower energies - but as in the case of the increase in tank size, the higher signal rate enables more aggressive background rejection and so overall sensitivity is increased. This results in a decrease in dwell time with the addition of WbLS.

\begin{table}[htb]
    \centering
    \caption{Optimised cuts in the 16~m (22~m) Gd-water detector for anomaly detection of all signal combinations, where the cuts are described in section~\ref{cobraa:optimisation}.}
    \begin{ruledtabular}
    \begin{tabular}{lcccccc}
     Signal            & Fiducial (m from PMTs) & $E_{p,min}$ (hits) & $E_{d,min}$ (hits) & $dT_{max}$ ($\mu$s)  & $E_{p,max}$ (hits) & $g_{min}$   \\
     \hline
    Hartlepool 1\&2    &   1.2 (1.6)   &  9 (9)   &    9 (9)     & 120 (150)   &   35 (45)   & 0.5 (0.6)\\
    Hartlepool 2       &   0.9 (0.9)   &  9 (10)  &    9 (14)    & 120 (120)   &  35 (40)    & 0.2 (0.6) \\
    Heysham            &   1.7 (2.2)   & 9 (9)    &   10 (11)    & 200 (200)   & 20 (30)     & 0.7 (0.6)\\
    Heysham Torness    &   2.7 (2.2)   & 9 (9)    &    9 (11)    & 200 (200)   & 25 (35)     & 0.6 (0.6)\\
    Heysham 2          &   2.5 (2.2)   & 9 (9)    &   11 (11)    & 200 (200)   & 20 (30)     & 0.6 (0.6)\\
    \end{tabular}
    \end{ruledtabular}
    \label{tab:cobraa_cuts_water}
\end{table}

\begin{table}[htb]
    \centering
    \caption{Optimised cuts in the 16~m (22~m) Gd-WbLS detector for anomaly detection of all signal combinations, where the cuts are described in section~\ref{cobraa:optimisation}.}
    \begin{ruledtabular}
    \begin{tabular}{lcccccc}
     Signal            & Fiducial (m from PMTs) & $E_{p,min}$ (hits) & $E_{d,min}$ (hits) & $dT_{max}$ ($\mu$s)  & $E_{p,max}$ (hits) & $g_{min}$   \\
     \hline
    Hartlepool 1\&2    &  0.9 (1.5)    &  15 (17)    &   23 (15)    & 190 (260)  &   60 (75)  &  0.4 (0.4) \\
    Hartlepool 2       &  0.6 (0.7)    &  21 (15)    &   27 (27)    & 210 (250)  &  60 (60)   &  0.4 (0.4)\\
    Heysham            &  1.1 (1.6)    &  27 (21)   &   13 (13)     & 210 (230) &  55 (55)    &  0.4 (0.4) \\
    Heysham Torness    &  1.1 (1.6)    &  27 (27)   &   13 (13)     & 210 (230) &   55 (55)   &  0.4 (0.3)  \\
    Heysham 2          &  1.3 (1.6)    &  27 (25)   &   13 (11)     & 240 (230) &   50 (55)   &  0.4 (0.4) \\
    \end{tabular}
    \end{ruledtabular}
    \label{tab:cobraa_cuts_wbls}
\end{table}

\subsection{LEARN}
\label{sec:Learn}

The LEARN (Likelihood Event Analysis of Reactor Neutrinos) analysis~\cite{Learn2021} is based on different principles to Cobraa. LEARN utilizes a likelihood ratio test, supervised machine learning using AdaBoost~\cite{scikit-learn}, and energy analogue cuts to form an analysis chain.

There are several key steps involved with the LEARN pathway. While the Monte Carlo generation is identical to that used in Cobraa, the position reconstruction of the events differs. Here, the BONSAI event-by-event algorithm that CoRe is built upon is used. Following this, a multiplicity cut as described in Section~\ref{coincidence:cosmics} is applied before the main analysis. This removes chains of events where more than two particles are detected to reduce the fast-neutron background. After the multiplicity cut, a likelihood ratio test removes uncorrelated single events coming from radioactivity that have the potential to produce accidental coincidences. 
Events that are deemed to be in a pair by the likelihood test are passed into a machine learning algorithm which is designed to identify and remove remaining fast neutrons. After the suppression of fast neutrons, energy cuts are used to reduce radionuclide and low-energy backgrounds before an analytical post-muon veto is applied (Section~\ref{sec:muon-veto}).

\subsubsection{Likelihood ratio test statistic}
\label{subsec:LRT}

The likelihood component of the analysis chain is used to handle accidental coincidences from radioactive backgrounds occurring naturally in the detector and surrounding environment. The majority of these events originate from the PMTs and so occur close to the edge of the inner detector volume. As these events are uncorrelated to any other events, have very low energies and occur near the edge of the inner volume, they can be distinguished from correlated events.

The Monte Carlo is split into two parts, a training data set and an evaluation data set. Probability density functions (PDFs) are created using the training Monte Carlo for each signal and background source for each parameter, before being used to create signal and background likelihoods. LEARN uses a likelihood ratio test based on five parameters to separate out uncorrelated events. The parameters are:

\begin{enumerate}
\item Number of PMTs hit within 100 ns of the peak of light from an event (n100),
\item Number of PMTs hit within 100 ns of the peak of light from an event for the previous event,
\item Time between two consecutive events,
\item Distance between two consecutive events,
\item Distance from PMT support structure.
\end{enumerate}

\begin{figure}[htb]
    \centering
        \includegraphics[width=86mm]{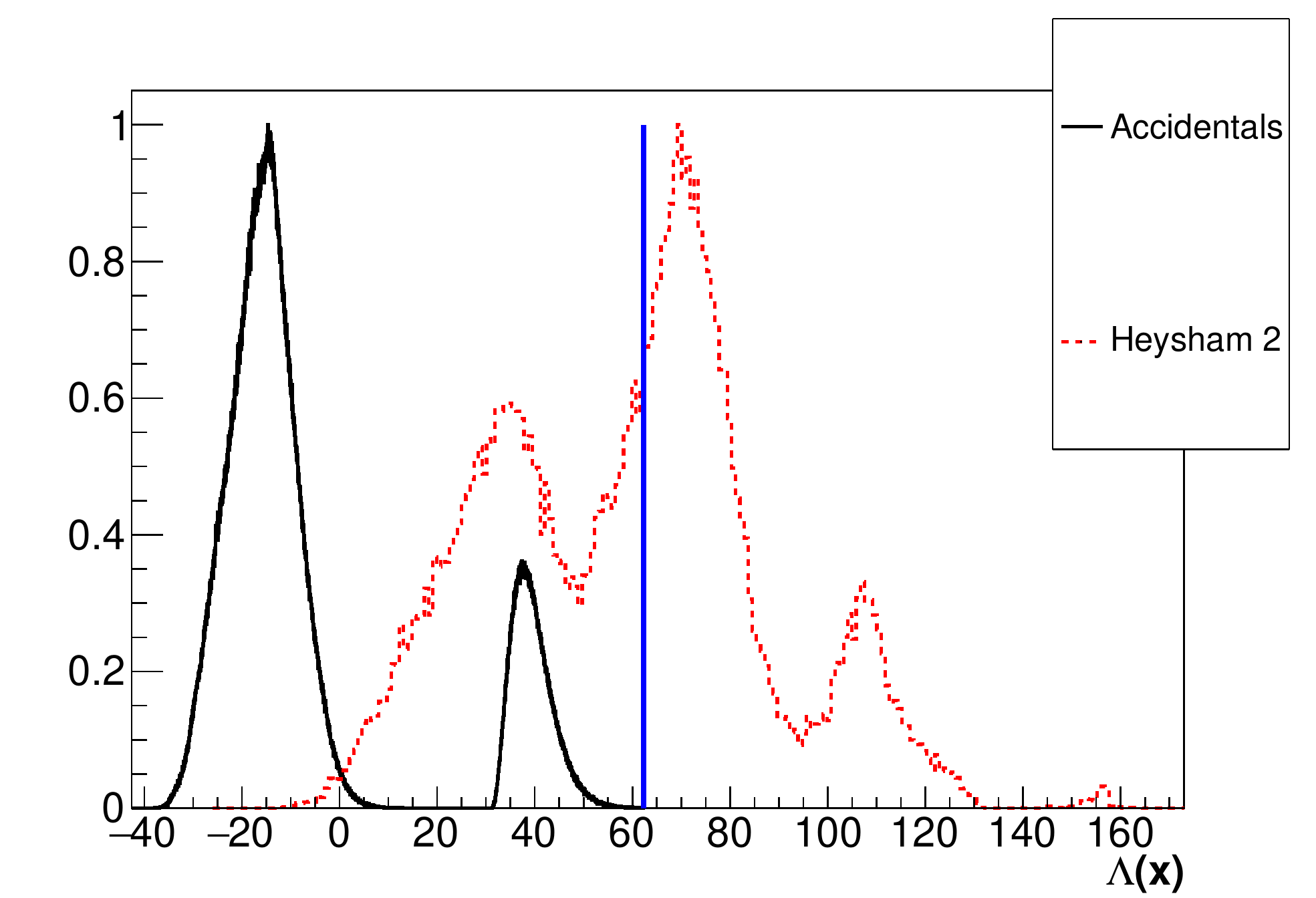}
    \caption{The test statistic defined in Equation~\ref{eq:Wilks} for the uncorrelated events (solid black) and IBD from Heysham 2 (dashed red) in the 22~m Gd-WbLS configuration. The blue vertical line is the upper boundary of the distribution for the uncorrelated events.}
        \label{fig:like}
\end{figure}

The signal and background likelihood values of an event are determined, and a test statistic is defined as
\begin{equation}
    \Lambda(x) = -2\ln\bigg(\frac{L(x|\theta_b)}{L(x|\theta_s)}\bigg)
    \label{eq:Wilks}
\end{equation}
where $\theta_s$ and $\theta_b$ are the probability distributions of the parameters for the correlated and uncorrelated event components respectively, and $L$ is the likelihood. The statistic is used to discriminate between the correlated and uncorrelated components using by tuning a cut to match the upper bound of the distribution produced by uncorrelated events. This is to allow the removal of all uncorrelated events while keeping the maximum number of signal events possible. Evaluation data are then passed through the PDFs to determine their test statistic, and events are kept for further analysis if they pass the previously optimized cut. Some events may be determined to have a non-zero probability of being signal and a zero probability of being background from the PDFs, but be rejected by the optimized cut. In this situation, the events are kept. An example of the distribution of the test statistic for IBD events from Heysham 2 and uncorrelated events, with the cut made to remove uncorrelated events, is displayed in Fig.~\ref{fig:like}.

The nature of Fig~\ref{fig:like}, with its multiple peaks, is due to the event types being compared. Correlated and uncorrelated events are passed through both their own PDFs and the other event type's PDFs. The uncorrelated events are single events and create smooth distributions, whereas the correlated event has two events (the positron and neutron) within its distributions. This results in peaks caused by the differing event types within the correlated event likelihood distributions. When the likelihoods are divided, these peaks propagate through and give the distributions shown in Fig.~\ref{fig:like}.

Since the null hypothesis (uncorrelated background) is one, continuous component, whereas the alternative hypothesis (coincident signal pair) is made up of two distinct components, there are times when one or other of the components are not defined for the background, leading to the two distinct peaks in the test statistic for the uncorrelated backgrounds.

\begin{figure}[htb]
    \centering
        \includegraphics[width=86mm]{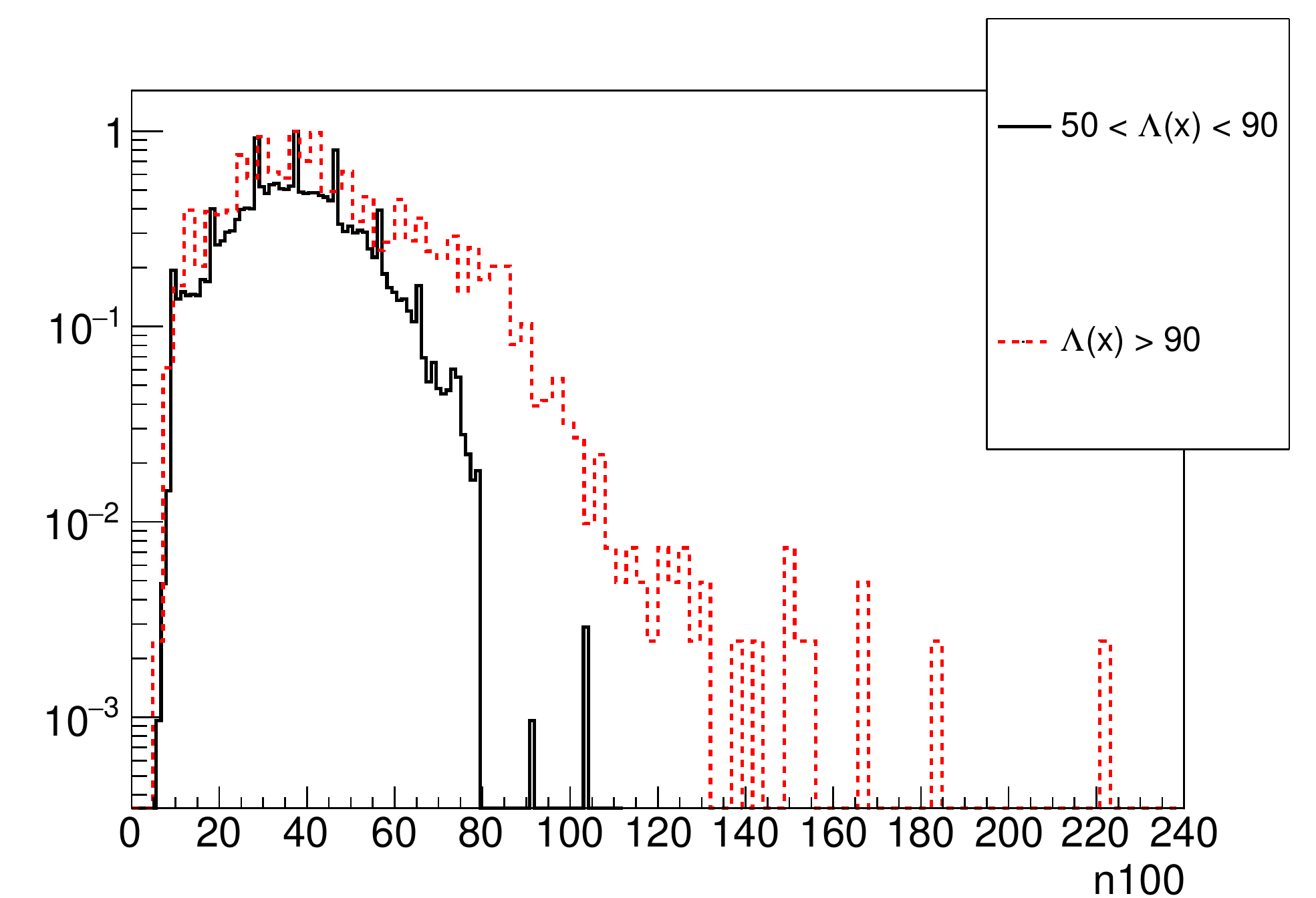}
    \caption{The number of PMT hits within 100 ns of the event (n100) for the positrons in the likelihood ratio peaks in Fig.~\ref{fig:like} corresponding to 50 $<$ $\Lambda(x)$ $<$ 90 (solid black) and $\Lambda(x)$ $>$ 90 (dashed red).}
        \label{fig:n100-peaks-2-3}
\end{figure}

The three peaks in the Heysham 2 distribution corresponds correspond to events with different properties. The first peak, with a value of $\Lambda(x)$ $<$ 50, is mostly neutrons in IBD pairs. There is a tail that extends across the majority of the positive $\Lambda(x)$ distribution. The second peak, with 50 $<$ $\Lambda(x)$ $<$ 90, consists mostly of positrons in an IBD pair where both events trigger. As such, the first two peaks are closely linked. The third peak is generally positrons with a much higher number of PMT hits within 100~ns. This is because they deposit a higher amount of energy and so produce more light. The n100 distributions of the second and third peak are shown in Fig.~\ref{fig:n100-peaks-2-3}. These positrons are also less likely to be followed by a neutron that triggers the detector. They pass the optimized cut as they have a much higher energy than single radioactive events and occur much nearer the center of the detector, therefore they do not need the correlated event to be discriminated from radioactive backgrounds.

\subsubsection{Machine learning}
\label{subsec:ML}

Once coincidences are evaluated and uncorrelated backgrounds removed, the data are passed into a machine learning model to remove fast neutrons. The model used is a boosted decision tree called AdaBoost~\cite{scikit-learn}, in which data are passed through a ``forest" of short decision trees and events classified by their features. Misclassified data are given an increased weighting on each iteration (boosted), allowing harder-to-classify events to be prioritized. A confidence score can also be given and is based on how many times an event is classified as signal or background as it passes through the forest.

\begin{figure}[htb]
 \begin{subfigure}[b]{8.6cm}
  \includegraphics[width=\textwidth,height=7cm]{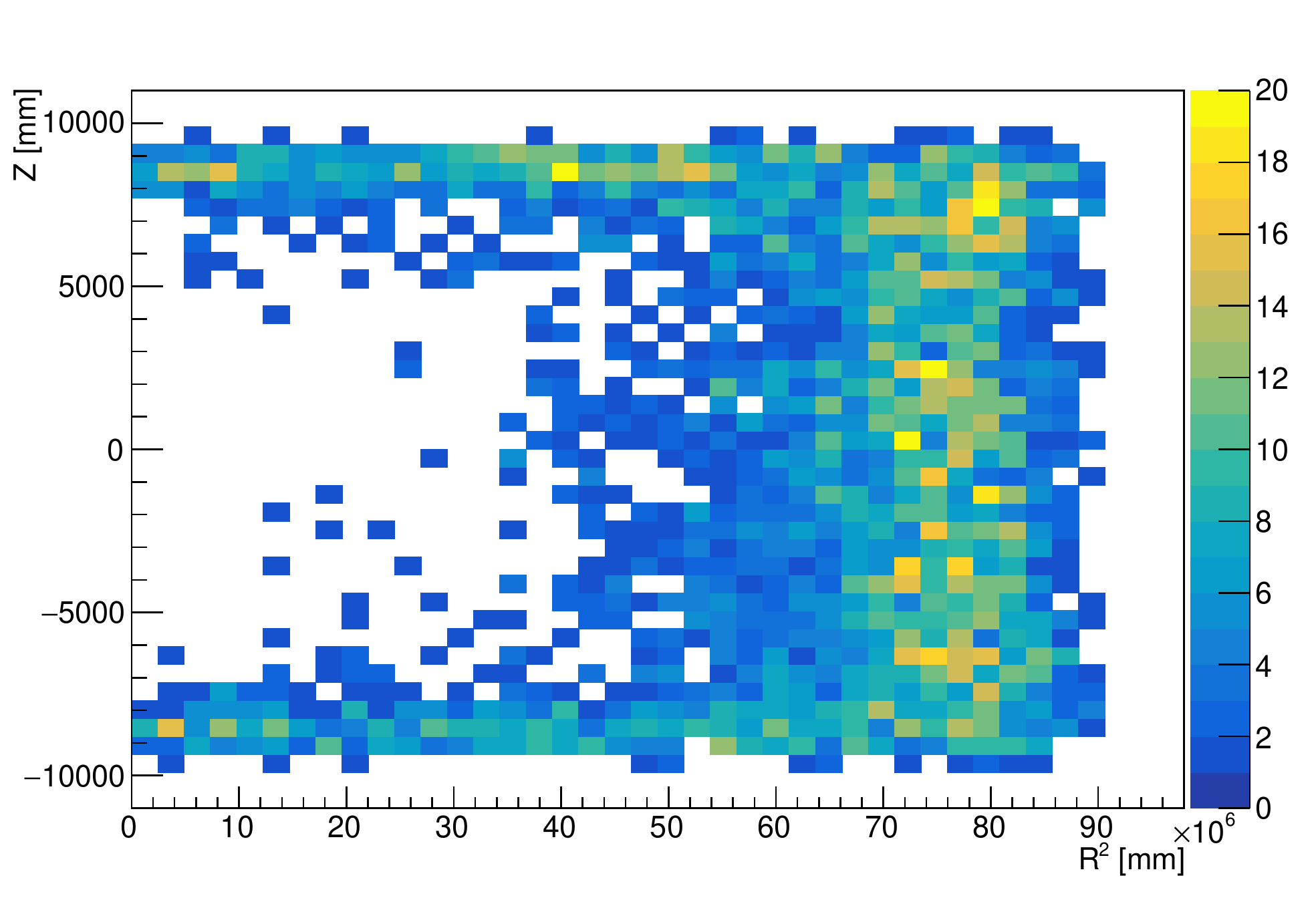}
  \caption{}
  \label{fig:fnpos}
 \end{subfigure}
 \hfill
 \begin{subfigure}[b]{8.6cm}
 \includegraphics[width=\textwidth,height=7cm]{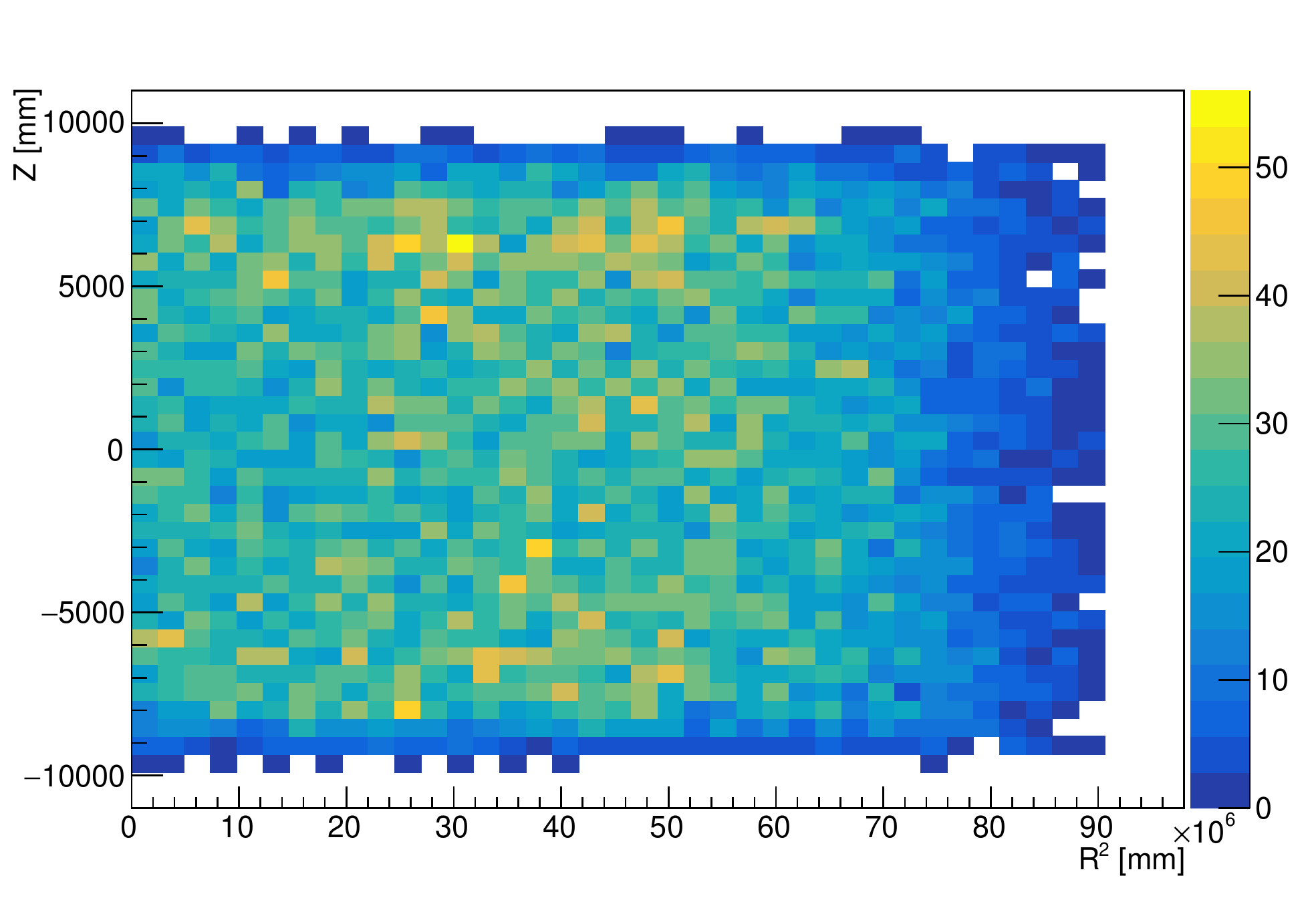}
 \caption{}
 \label{fig:heypos}
 \end{subfigure}
 \caption{Reconstructed event position in 22~m tank filled with Gd-WbLS for fast neutrons (a) and IBD events (b). This is used in the machine learning model to help discriminate between the signal and fast neutron background.}
 \label{fig:pos}
\end{figure}

The model is trained on the training Monte Carlo data set used in the likelihood analysis step, with the final classification made on the pairs of events in the evaluation data kept by the likelihood ratio test. The model trains on several parameters, most of which are the same as used in the likelihood ratio test, with additional parameters taking into account vertex reconstruction quality and more detail on the position of the event. The difference in the distribution of event positions is shown in Fig.~\ref{fig:pos} and shows good discrimination between fast neutrons and IBD events, since IBD events happen throughout the detector whereas fast neutrons are concentrated near the edge. As such, additional position information allows a better classification of events.

\begin{figure}[htb]
 \begin{subfigure}[b]{8.6cm}
  \includegraphics[width=\textwidth]{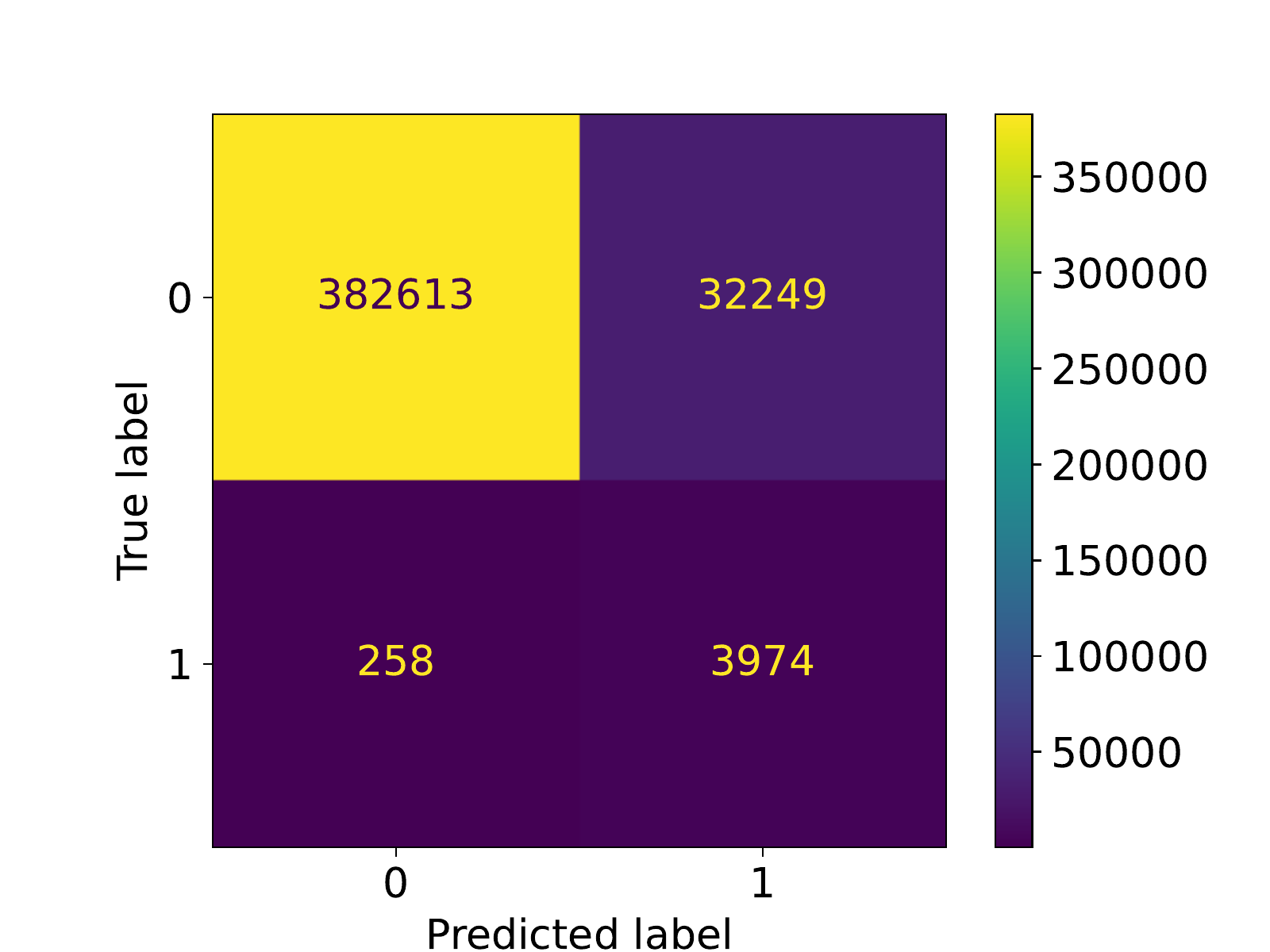}
  \caption{}
  \label{fig:neutronCM}
 \end{subfigure}
 \hfill
 \begin{subfigure}[b]{8.6cm}
 \includegraphics[width=\textwidth]{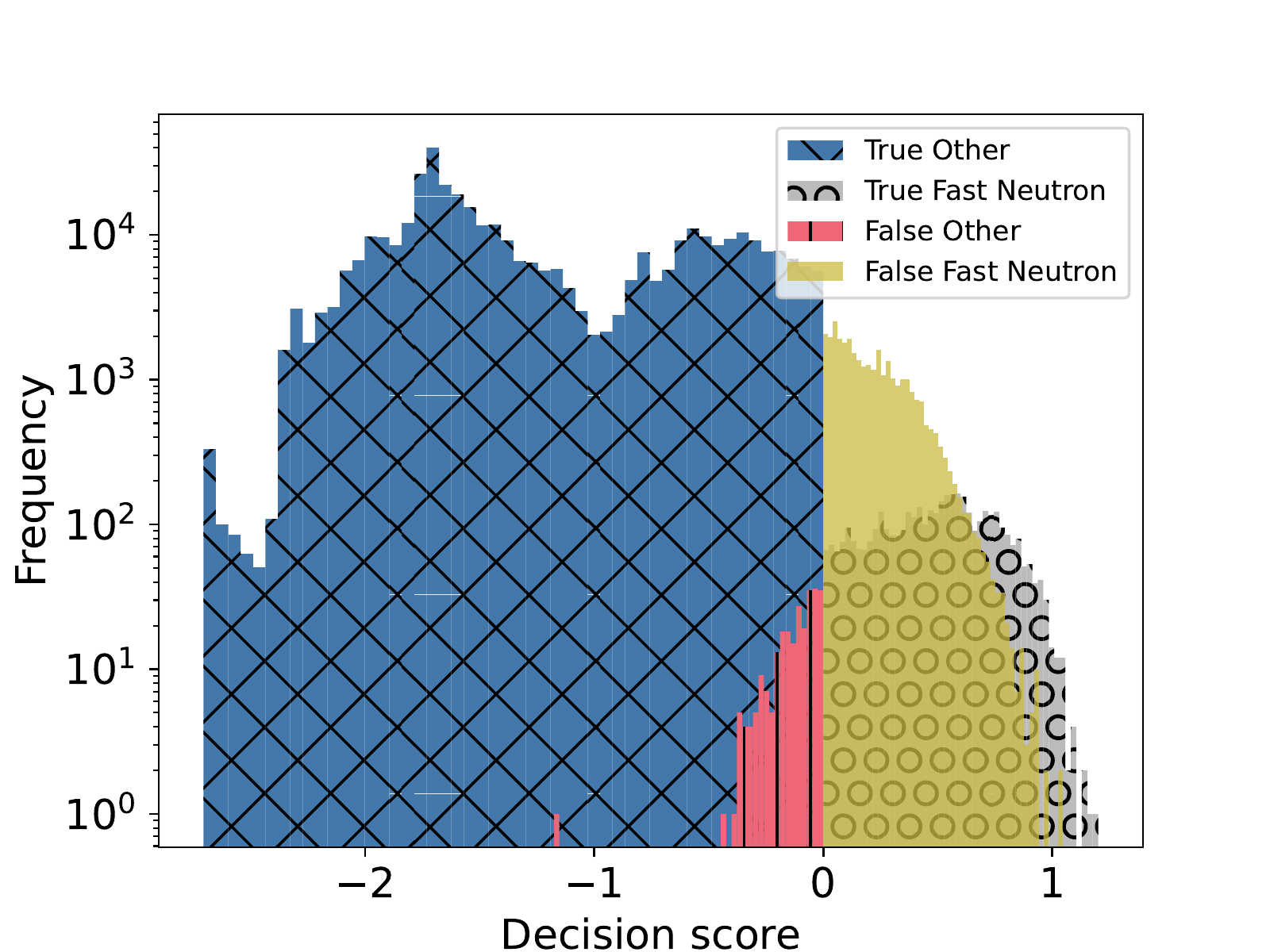}
 \caption{}
 \label{fig:neutronScores}
 \end{subfigure}
    \caption{The classification of fast neutrons in the 22~m tank filled with Gd-WbLS with Heysham 2 as the target reactor for the evaluation data set. In the confusion matrix (a), fast neutrons are marked as 1 and all other sources are marked as 0. In the confidence scores (b), events labeled `other' include any that are not fast neutrons. Only 286 out of 26096 Heysham 2 events are rejected by the classifier.}
\end{figure}

The model is applied to look for a specific background source, fast neutrons, and not a signal source as is often the case. This is done to harness the differences in properties between the fast neutrons that reach the inner detector volume and other correlated sources. Event pairs tagged as fast neutrons are removed from the data. This is done with $>$ 94\% efficiency and keeps $\sim$ 99\% of reactor IBD, which is shown by the confusion matrix for Heysham 2 in the 22~m Gd-WbLS configuration in Fig.~\ref{fig:neutronCM}, with the decision score associated with this classification shown in Fig.~\ref{fig:neutronScores}. The majority of the falsely rejected events are $^9$Li, which further reduces background. This is likely due to the higher energy of the prompt $^9$Li electron, which produces light more comparable to a fast neutron than an IBD positron. A comparison between the kinetic energy of the prompt particles from $^9$Li and IBD is presented in Fig.~\ref{fig:max_ep_cut_li9} in Section~\ref{sec:sims}.


\subsubsection{Energy Cuts}
\label{subsec:Energy}

Following the removal of uncorrelated single events and the suppression of fast neutrons, there are still several types of pairs of events. These originate from reactor antineutrinos, muon-induced radionuclides, geoneutrinos and the remaining fast neutrons. While different reactors cannot be distinguished in these detector configurations, except where oscillations cause the signal and reactor IBD background spectra to diverge significantly, the other background types can be further reduced due to their difference in energy to reactor IBD.

Geoneutrinos tend to have a lower energy than reactor antineutrinos (Fig.~\ref{fig:antinu_backgrounds}), while fast neutrons and $^9$Li can have a higher prompt energy than the positron produced via IBD from reactor antineutrinos. Energy cuts in terms of an energy analogue can be optimized to harness these differences. The energy analogue (Section~\ref{cobraa:optimisation}) chosen here is n100 to make the most of the slower scintillation light by allowing 100 ns of light from an event to be used. Threshold and maximum values for n100 are optimized for the prompt signal in a pair, and the pair is kept if the prompt event passes both cuts. Minimum and maximum n100 values for the delayed signal are also optimized, with the pair being removed if the delayed event energy is not within these two values. For some configurations, cuts on the delayed signal energy were not required. Table~\ref{tab:learn_cuts_wbls} shows the optimized cuts for the Gd-WbLS configurations. 

\begin{table}[htb]
    \centering
    \caption{Optimised energy cuts, as described in Section~\ref{subsec:Energy}, in the 16~m (22~m) Gd-WbLS detector for anomaly detection of all signal combinations using LEARN. Cuts on the delayed signal energy are not required in all cases.}
    \begin{ruledtabular}
    \begin{tabular}{lcccc}
     Signal            & $E_{p,min}$ (hits) & $E_{p,max}$ (hits) & $E_{d,min}$ (hits) & $E_{d,max}$ (hits) \\
     \hline
    Hartlepool 1\&2    &  10 (0)    &  90 (102)   &   N/A (N/A)     & N/A (N/A) \\
    Hartlepool 1       &  0 (0)    &  110 (104)   &   N/A (N/A)     & N/A (N/A) \\
    Heysham            &  20 (22)    &  70 (70)   &   80 (N/A)     & 190 (N/A) \\
    Heysham 2 + Torness    &  20 (24)    &  80 (85)   &   80 (N/A)     & 250 (N/A) \\
    Heysham 2          &  20 (24)    &  80 (64)   &   80 (N/A)     & 250 (N/A) \\
    \end{tabular}
    \end{ruledtabular}
    \label{tab:learn_cuts_wbls}
\end{table}

Following the energy cuts, an analytical post-muon veto is applied as detailed in Section~\ref{sec:muon-veto} to further reduce event rates from muon-induced backgrounds.

\subsection{Analytical Post-muon veto}
\label{sec:muon-veto}

The final stage before calculation of the dwell time in both the analyses is to apply an analytical post-muon veto, which rejects many of the remaining radionuclide backgrounds.

Correlations between radionuclide-like events and a muon track can be used to reduce the radionuclide backgrounds. Backgrounds from long-lived radionuclides with half lives on the order of seconds would survive a short time veto after a muon event. With effective muon tracking, applying a longer veto within a limited transverse distance from a muon track can help to reduce these backgrounds without introducing excessive dead time in the detector. The fraction of radionuclide activity remaining after a veto time $t_{veto}$ with a fractional muon detection efficiency $\epsilon$ is given by:

\begin{equation}
    \rm R_{iso,}(t_{veto}) = \bigg(1-\epsilon  + \epsilon \frac{\int_{t_{veto}}^\infty e^{-ln(2)t/t_{1/2}}dt}{\int_0^\infty e^{-ln(2)t/t_{1/2}}dt}\bigg) ~R_{iso,tot}.
\end{equation}

With an active veto, the muon-detection efficiency might reasonably be expected to be as high as 99.9\%, allowing for electronics issues. With a passive buffer rather than an active veto, the muon-detection efficiency in the fiducial volume might be assumed to be the same. However, making some correction for muons which pass through the veto but induce spallation in the fiducial volume, a conservative muon-detection efficiency of 95\% for a passive buffer is assumed and the fractional reductions in radionuclide rates with a 1~s veto are:
\begin{align*}
\rm R_{^9Li}(1~s ~veto) &= \rm 0.069~R_{^9Li,tot}\\
\rm R_{^{17}N}(1~s ~veto) &= \rm 0.85~R_{^{17}N,tot}\\
\end{align*}

Making the reasonable assumption that it will be possible to track the muons passing through the fiducial volume, the 1~s time veto is assumed to be imposed only over a limited transverse distance from the muon track and thus incurs negligible detector dead time. 

\section{Results and validation}
\label{sec:results}

Since this study relies on simulation, an evaluation of the similarities and differences between the two independent reconstruction-analysis pathways has been used to validate the results. Results from the two analysis paths are presented and compared where appropriate in Section~\ref{subsec:cobra-res}. An extension to more distant reactor signals is presented in Section~\ref{subsec:PWRresults} and implications for all results are discussed in Section~\ref{subsec:disc}.

\subsection{Results for the AGR reactors}
\label{subsec:cobra-res}

The CoRe reconstruction and Cobraa analysis were applied to all signal combinations and all detector configurations. Both detector sizes were also analyzed with LEARN, but only with the Gd-WbLS fill. A direct comparison of Cobraa and LEARN results can be made for the 16~m and 22~m Gd-WbLS detector configurations, for five different AGR reactor signals.

Table~\ref{tab:cobraa_results} summarizes the results obtained for the AGR signals in terms of the number of days of observation required for a 3$\sigma$ rejection of the no-reactor hypothesis for each detector configuration studied. The sensitivity to the reactor signal increases and the dwell time comes down with the addition of WbLS and with the larger detector size in all scenarios. None of the configurations were able to give reasonable sensitivity, within practical limits, to the single-site Torness signal.

\begin{table*}[htb]
    \caption[Results summary - anomaly detection]{Results summary for anomaly detection - dwell time in days for rejection of the background-only hypothesis to 3$\sigma$ significance assuming normal reactor operation. The percentage difference and difference in days between Cobraa and LEARN are shown in brackets where relevant.}
    \begin{ruledtabular}
    \begin{tabular}{cccccccc}
   Analysis   &   {Detector}       & {Hartlepool} & {Hartlepool 1}    & {Heysham}  & {Heysham 2}      & {Heysham 2}  \\
              &          & {1 \& 2}     &                 & {1 \& 2}   & {+ Torness}      &            \\
         \hline
   Cobraa &    16~m Gd-H$_2$O &   12       & 61            &  2327       &  3488          & 8739   \\
   Cobraa &      16~m Gd-WbLS   &  7         & 35            & 738         & 1022           & 3008   \\
   LEARN  &      16~m Gd-WbLS   & 5         & 24            & 1017          & 963           & 2968 \\ 
                       &        & (29\%, 2) & (31\%, 9)     & (27\%, 279)   & (6\%, 59)     & (1\%, 40) \\ 
   Cobraa &      22~m Gd-H$_2$O &   3        &  11           & 241          & 232  & 985    \\  
   Cobraa &      22~m Gd-WbLS   & 2          & 8             & 152            & 192  &  647    \\
    LEARN &     22~m Gd-WbLS    & 2         & 9             & 196           & 164       & 577 \\
                        &        & (0) & (11\%, 1) & (22\%, 44) & (15\%, 28) & (11\%, 70) \\ 
    \end{tabular}
    \end{ruledtabular}
    \label{tab:cobraa_results}
\end{table*}

The results for reactor measurement are shown in Table~\ref{tab:cobraa_reactor_measurement} for the 22~m Gd-WbLS configuration. It would take an additional $\sim160$~days ($\sim140$~days) - around 25\% longer - to make a reactor measurement of Heysham 2 after an initial anomaly detection according to Cobraa (LEARN). Cobraa found that measurement of the stronger Heysham 1~\&~2 and Heysham~2~+~Torness signals would take  respectively $\sim16$\% and $\sim7$\% longer than an anomaly measurement. LEARN found that reactor measurement would take $\sim35$\% longer for the same signal combinations.

\begin{table}[htb]
    \caption{Results summary for reactor measurement - dwell time in days for confirmation of the signal-plus-background hypothesis to 3$\sigma$ significance under normal reactor operation.}
    \begin{ruledtabular}
    \begin{tabular}{ccccc}
 Analysis       & { Detector} &  { Heysham} & { Heysham 2}   & { Heysham 2}\\
                &         & { 1 \& 2}  & { + Torness}  &  \\
         \hline
Cobraa & 22~m Gd-WbLS & 176 & 206 & 808 \\ 
LEARN & 22~m Gd-WbLS & 263 & 221 & 715 \\
     &      &  (33\%, 87) & (7\%, 15) & (12\%, 93) \\
    \end{tabular}
    \end{ruledtabular}
    \label{tab:cobraa_reactor_measurement}
\end{table}

The two analysis paths agree for all signal/detector configurations to between the 0\% and 31\% level. In most cases, the small dwell times for the Hartlepool 1~\&~2 and Hartlepool 1 signals results in large percentage differences between the two analyses where in fact the dwell times differ by only a matter of days. This is made more significant by the rounding of the dwell times to the nearest day. For the more distant reactors, the dwell times differ by between around one and two months, with the exception of the dwell times for anomaly detection of the Heysham~1~\&~2 signal with the 16~m Gd-WbLS detector. 

In the absence of data, the general agreement of the two analysis pathways helps to build confidence in the results. For this reason, it is important to compare the results and understand why the results diverge in places. The differences between the results of the analyses arise from the differing data reduction methods in combination with the optimization. Cobraa places more emphasis on background suppression for distant reactors, whereas LEARN places the emphasis on signal maximization, as demonstrated by Table~\ref{tab:rates_breakdown}, which shows the post-optimization signal and background rates for the 22~m Gd-WbLS detector from both analyses for anomaly detection. This is highlighted by the fact that the two analyses show opposite trends for the Heysham~1~\&~2 and Heysham~2~\&~Torness signals in all but one case. In the case of the Heysham~1~\&~2 signal, the signal spectrum has a peak at a higher energy, which allows a more aggressive cuts-based rejection of backgrounds than for the Heysham~2~\&~Torness signal. Dwell times from Cobraa, which optimizes on a signal-by-signal basis, are therefore shorter - despite the larger raw backgrounds due to the higher contribution from reactor IBD backgrounds. In contrast, dwell times from LEARN are longer for the Heysham~1~\&~2 signal but LEARN is able to maximize the lower signal from Heysham~2~\&~Torness as it has less background to contend with, resulting in shorter dwell times for the dual-site signal.

For anomaly detection, the 16~m detector with a Gd-H$_2$O fill is statistics limited. The optimized rates for each of the signals are given for the 22~m detector configurations in Table~\ref{tab:rates_breakdown} and for all configurations for the Heysham 2 signal in Table~\ref{tab:optimal_rates}. The 22~m Gd-H$_2$O detector becomes statistics limited for the single-site Heysham~2 signal. The detectors with the Gd-WbLS fill tend to be more systematics limited, particularly the 22~m configuration, and reducing the systematics on the backgrounds could bring the dwell time down. Reduction of systematic uncertainties on the backgrounds could be achieved through sideband analysis or through background measurements.

\begin{table*}[htb]

    \caption{Representative rates per day from optimization in the 22~m detector with Gd-WbLS fill. Accidental coincidences and geoneutrino IBDs are omitted as they are sub-dominant backgrounds.}
    \begin{ruledtabular}
    
    \begin{tabular}{lccccccc}
    
      Signal combination          &  Analysis &  s    & $\Sigma$ b$_i$ & $\rm{^9Li}$ \& $\rm{^{17}N}$  & Fast n & IBD${_{reactor}}$ &  ${\Sigma} {{\sigma}_{b_i}}$   \\
     \hline
     Hartlepool 1   & Cobraa & 2.2 & 3.2 & 0.27 & 0.54 & 2.4 & 0.20 \\
                    & LEARN  &  1.9   & 2.7 & 0.29 & 0.27 & 2.0 & 0.079 \\
                    \\
     Heysham 1 \& 2 & Cobraa & 0.18 & 0.48 & 0.13 & 0.044 & 0.30 & 0.021\\
                    & LEARN  & 0.19 &  0.65  & 0.19 & 0.059 & 0.35 & 0.025  \\
                    \\
     Heysham \& Torness & Cobraa & 0.13 & 0.33 &  0.085 & 0.038 & 0.21 &  0.016\\
                    & LEARN  & 0.22 & 0.75  & 0.25  & 0.058 & 0.37&0.029\\
                    \\
     Heysham 2  & Cobraa & 0.091 & 0.39 & 0.097&0.034   & 0.26& 0.018\\  
                    & LEARN  &  0.13   & 0.71  & 0.21 & 0.047 & 0.39 & 0.024 \\
     
     \end{tabular}
     \end{ruledtabular}
    \label{tab:rates_breakdown}
\end{table*}

\begin{table*}[htb]
    \caption{Representative rates per day from optimization in all detector configurations for the Heysham 2 signal. Accidental coincidences and geoneutrino IBDs are omitted as they are sub-dominant backgrounds.}
    \begin{ruledtabular}
    \begin{tabular}{lccccccc}

     Detector         & Analysis &  s    & $\Sigma$ b$_i$ & $\rm{^9Li}$ \& $\rm{^{17}N}$  & Fast n & IBD${_{reactor}}$ &  ${\Sigma} {{\sigma}_{b_i}}$   \\
     \hline
     16~m Gd-H$_2$O   & Cobraa & 0.0039 & 0.015 & 0.00048 & \num{2.7e-9}& 0.0010 & 0.00061\\
                    \\
     16~m Gd-WbLS & Cobraa & 0.017& 0.068 & 0.020 & 0.0035&0.044 &0.0028 \\
                    & LEARN  & 0.024& 0.14&0.044&0.010&0.079&0.0043\\
                    \\
     22~m GdH$_2$O & Cobraa & 0.059 & 0.26 & 0.069 & 0.011 & 0.18 &0.011\\

                    \\
     22~m Gd-WbLS & Cobraa & 0.091 & 0.39 & 0.097&0.034   & 0.26& 0.018\\  
                    & LEARN  &  0.13   & 0.71  & 0.21 & 0.047 & 0.39 & 0.024 \\
     \end{tabular}
     \end{ruledtabular}
    \label{tab:optimal_rates}
\end{table*}
 
The differences between the analyses make the discrepancies between the dwell times for 3$\sigma$ reactor measurement (Table~\ref{tab:cobraa_reactor_measurement}) larger for the Heysham-only results. There is just a fifteen-day (7\%) difference between the results for the Heysham~2~\&~Torness signal.

\subsection{Results for distant PWR reactors}
\label{subsec:PWRresults}

After 2028, all of the cores at Hartlepool, Heysham and Torness will be decommissioned according to the current schedule. Beyond this time, the next nearest reactors to Boulby Mine will be the PWR reactors at Sizewell~B (306~km) and Hinkley Point~C (404~km) in the UK and Gravelines (441~km) in France, according to current schedules~\cite{EDFEnergy,ASN}. Evaluation of the sensitivity of the four detector geometries was extended to combinations of signals from these three nuclear power stations as detailed in Table~\ref{tab:PWRs}. 

The projected power of the two cores at Hinkley Point C was calculated for the purposes of this study from the 3.260~GW$_e$ power generation capacity projected by EDF energy~\cite{EDFhinkley} and the relation between thermal power and electricity generation from~\cite{Imeche}:
\begin{equation*}
    \rm 3.260 ~GW_e \left(\frac{4.3~GW_{th} }{1.6~GW_e}\right) = 8.76 ~GW_{th}
\end{equation*}
and the detectable signal spectrum and strength at Boulby was calculated using the custom core feature from~\cite{Dye2021}.

\begin{table*}[htb]
    \caption{Combinations of PWR reactor signals evaluated, along with their reactor backgrounds. Reactors marked black are included in the signal and reactors marked grey are included in the background. Signal and background rates shown in NIU.}
    \begin{ruledtabular}
    \begin{tabular}{lrrrr}
    \bf Signal combination & Sizewell B    & Hinkley Point C     & Gravelines       &  World \\
    \hline
    \bf Sizewell-Hinkley-Gravelines  
    &\multicolumn{3}{r}{\cellcolor{black}{\textcolor{white}{\bf 37~}}} & {\cellcolor{DGray}{\bf 84 }}   \\
    \hline
    \bf Sizewell-Hinkley  &\multicolumn{2}{r}{\cellcolor{black}{\textcolor{white}{\bf 22 }}} & \multicolumn{2}{r}{\cellcolor{DGray}\bf{99~~}}   \\
    \hline
    \bf Hinkley-Gravelines & \multicolumn{1}{r}{\cellcolor{DGray}{\bf  }} & \multicolumn{2}{r}{\cellcolor{black}{\textcolor{white}{\bf 29 }}} & \multicolumn{1}{r}{\cellcolor{DGray}{\bf 92\footnote{Total background including Sizewell B complex.}}}  \\ 
    \end{tabular}
    \end{ruledtabular}
    \label{tab:PWRs}
\end{table*}

Results for the anomaly detection of the three PWR signal combinations are shown in Table~\ref{tab:compare-PWRresults} for the 22~m Gd-WbLS detector. Only this configuration was sensitive to these cores within a practically reasonable time.

Both analyses agree to 5\% that a three-site detection of all cores at Sizewell, Hinkley and Gravelines would require just over two years with this detector configuration and the results for the Hinkley-Gravelines signal agreed to 0.2\%. There was a large disparity in days between the dwell times to anomaly detection of the Sizewell-Hinkley signal combination but, since even the most optimistic dwell time is $\sim16$ years and therefore well beyond any practical time frame, this does not detract from the overall agreement of the results. The dwell time calculation for anomaly detection did not converge for any of the three PWR nuclear power stations individually.

\begin{table}[htb]
    \caption{Comparison of Cobraa and LEARN results for anomaly detection - dwell time in days for rejection of the background-only hypothesis to 3$\sigma$ significance - in the 22~m Gd-WbLS detector for signal combinations of the PWR reactors at Sizewell~B (\textit{Size}), Hinkley Point~C (\textit{Hink}) and Gravelines (\textit{Grav}). The percentage difference and difference in days between Cobraa and LEARN are shown in brackets.}
    \begin{ruledtabular}
    \begin{tabular}{cccc}
    
         { Analysis} & { Size-Hink-Grav} & 
         {Hink-Grav}&{ Size-Hink}\\
         \hline
 Cobraa     & 785   & 1641  & 7077  \\
 LEARN      & 744  &   1638    &    5980    \\ 
            & (5\%, 41) & (0.2\%, 3) & (16\%, 1097)   \\ 
    \end{tabular}
    \end{ruledtabular}
    \label{tab:compare-PWRresults}
\end{table}

\subsection{Discussion}
\label{subsec:disc}

The 16~m detector configuration is well-suited to observing the antineutrino signal from the nearby reactors ($\rm <30~km$). However, it is not sufficient to make a single-site observation of a more distant target within a reasonable dwell time, as confirmed by both analyses.

The 22~m detector has significantly more potential. The Cobraa and LEARN analyses agree that a 22~m detector with Gd-WbLS would be able to make a single-site detection of the Heysham~1~\&~2 signal or a dual-site detection of the combined Heysham~2~\&~Torness signal within seven months, followed by a single-site detection of the signal from Heysham~2 about fifteen months later. 

\begin{figure}[htb]
    \includegraphics[width=86mm]{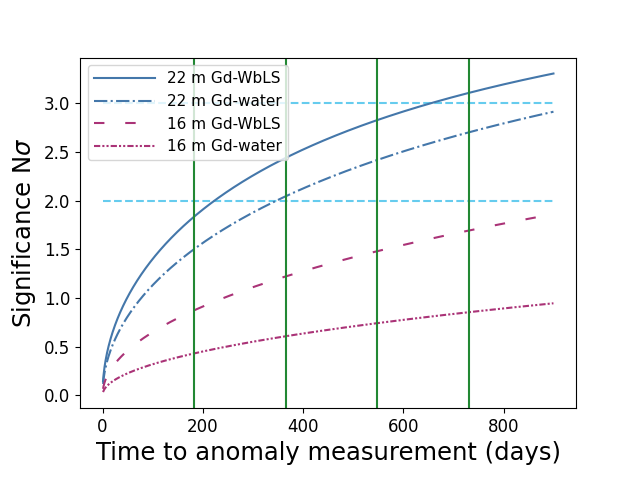}
    \caption[Sensitivity of all detector configurations to distant reactors]{The sensitivity of all detector configurations to anomaly detection of the Heysham 2 complex, using optimized signal and background rates for the most cautious (longer) dwell times. Shown as a function of observation time with vertical green lines marking 6, 12, 18 and 24 months from start of observation and horizontal blue lines marking 2$\sigma$ and 3$\sigma$ significance.
    }
    \label{fig:Heysham2_sensitivity}
\end{figure}

Fig.~\ref{fig:Heysham2_sensitivity} shows the Heysham 2 signal significance after 6, 12, 18 and 24 months of observation time for all detector configurations, using the signal, background and systematic uncertainties from Table~\ref{tab:optimal_rates} associated with the most cautious (longer) dwell times. Table~\ref{tab:22m_significance} shows the precise values of the significance at these observation times for all configurations. In the 22~m detector with a Gd-WbLS fill, a significance of 3.1 would be reached within the two-year period. A significance of 2.6 would be reached within the same time frame with the Gd-H$_2$O fill in the 22~m detector and it should be noted that this may be considered sufficient by the non-proliferation community. 

\begin{table}[htb]
     \caption[Anomaly detection significance for the Heysham 2 reactor complex]{Significance for anomaly detection of Heysham 2 at 6, 12, 18 and 24 months of observation time with all detector configurations, using optimized signal and background rates for the most cautious (longer) dwell times. The 22~m Gd-WbLS detector reaches 3$\sigma$ significance within two years.}
     \begin{ruledtabular}
    \begin{tabular}{ccccc}
    
     Detector &   6 months &  12 months &  18 months &  24 months  \\
     \hline
     16~m Gd-H$_2$O &  0.43     & 0.61     & 0.74     & 0.85  \\
     16~m Gd-WbLS   &  0.87   &  1.22    & 1.48  & 1.69 \\    
     22~m Gd-H$_2$O & 1.50      &  2.04    &  2.42    &  2.70 \\
     22~m Gd-WbLS   &  1.83     &  2.44    &  2.83     & \textbf{3.11}  \\
     
    \end{tabular}
    \end{ruledtabular}
    \label{tab:22m_significance}
\end{table}

A less stringent 2$\sigma$ detection of Heysham~2, which could be considered sufficient in combination with other monitoring methods, would be possible with the 22~m Gd-WbLS detector within around ten months according to the most cautious evaluation. Even with the most optimistic signal and background rates from Table~\ref{tab:optimal_rates}, the 16~m detector is unable to reach 2$\sigma$ in two years with either fill, suggesting it is too small for monitoring of more distant or smaller reactors, even in combination with other monitoring methods.

While Gd-WbLS is an innovative new fill material which shows much promise for remote reactor monitoring, Gd-H$_2$O has already been deployed in a large-scale water Cherenkov detector and associated technological risks are much lower. According to the Cobraa analysis, the 22~m detector should be able to complete a dual-site detection of the combined antineutrino signal from the AGR-2 reactors at Heysham 2 and Torness within a dwell time of less than 8 months, followed by a less stringent, 2$\sigma$ single-site detection of the antineutrino signal from Heysham 2 in just one year, with the confidence reaching 2.7$\sigma$ in two years. 

The signal-to-background ratio for the single-site Torness signal was so low as to make anomaly detection of the reactor impossible within practical limitations of the location and all detector geometries, even with the Gd-WbLS fill.

The consistency between analyses, and the ability to explain differences in terms of the analysis processes where differences occur, builds confidence in the accuracy of the results quoted. The largest percentage discrepancy for anomaly detection is around 30\% and occurs for the Hartlepool cores using the 16~m Gd-WbLS configuration. It should be noted however that these differences equate to just a matter of days due to the short dwell times involved. 

The largest discrepancy in days for anomaly detection, excluding the impracticable Sizewell-Hinkley result, occurs for the Heysham~1~\&~2 signal - again using the 16~m Gd-WbLS configuration. The generally larger discrepancy in the results for Heysham~1~\&~2 is borne out in the dwell times for reactor measurement. This trend is due to the differences in optimization between the two methods and points to a potential to drive down the dwell times still further with a combination of the two methods.

\section{Conclusions}
\label{sec:Conclusions}

In this paper, we have presented two analysis pathways to evaluate the sensitivity of a water-based Cherenkov detector to a nuclear reactor for remote monitoring and applied them to real reactor signals. The Cobraa pathway represents a complete simulation-reconstruction-analysis toolchain which incorporates a novel reconstruction and a multivariate cuts-based analysis. The reconstruction is specially adapted for Gd doping, a recent advance water-Cherenkov technology. The LEARN pathway combines a robust low-energy reconstruction with an innovative analysis incorporating a likelihood ratio test and machine learning for the essential rejection of the fast neutron background. This is the first detailed study of this kind, and as such the inclusion of two analysis pathways is intended to add confidence. With data as a reference, having two analysis pathways would facilitate an estimation of systematics. 

Four detector configurations and nine reactor signals were considered for this study - testament to the flexibility of the two analyses. Of the thirty-six combinations of detector configuration + reactor signal, dwell times to achieve 3$\sigma$ significance for anomaly detection have been presented for twenty-three of them. Dwell times to achieve 3$\sigma$ reactor measurement have also been presented for signal combinations from the more distant reactors at Heysham and Torness in the 22~m Gd-WbLS detector. The results for anomaly detection of the AGR signals, excluding the single-site Torness signal, are summarized in Table~\ref{tab:sum_res}. The results for the PWR signals in the 22~m Gd-WbLS detector are summarized in Table~\ref{tab:sum_res_pwr}.

\begin{table*}[htb]

    \caption{Summary of dwell times in days to achieve 3$\sigma$ significance for anomaly detection with different reactor targets for the AGR signals. For combinations where results exist for both analyses, the more conservative ({\it i.e.}, higher) dwell time value is used.}
    \begin{ruledtabular}
    \begin{tabular}{cccccc}
    
         {Detector} & { Hartlepool} & { Hartlepool 1} & { Heysham} & { Heysham 2}   & {Heysham 2}  \\
                         & { 1 \& 2}    &                    & {1 \& 2}  & {+ Torness}  &    \\
         \hline
16~m Gd-H$_2$O &  12      &      61       &   2327   &  3488  &  8739         \\
16~m Gd-WbLS   &  7      &       35       &   1017   &   1022    &  3008         \\ 
22~m Gd-H$_2$O &  3    &   11        & 241 & 232 & 985        \\
22~m Gd-WbLS   &  2    &    9          &  196     &192 & 647         \\
    \end{tabular}
    \end{ruledtabular}
    \label{tab:sum_res}
\end{table*}

\begin{table}[htb]

    \caption{Dwell times in days to achieve 3$\sigma$ significance for anomaly detection of the PWR reactors at Sizewell~B (\textit{Size}), Hinkley Point~C (\textit{Hink}) and Gravelines (\textit{Grav}). The more conservative ({\it i.e.}, higher) dwell time value from the two analyses is used.}
    \begin{ruledtabular}
    \begin{tabular}{cccc}
    
          Detector & { Size-Hink-Grav} & 
         {Hink-Grav}&{ Size-Hink}\\
        \hline 
        22~m Gd-WbLS & 785   & 1641  & 7077  \\
        
    \end{tabular}
    \end{ruledtabular}
    \label{tab:sum_res_pwr}
\end{table}

Set within the UK's nuclear landscape, the 22~m detector was shown to be the minimum required for sensitivity to the reactor signals of $\rm \sim3~GW_{th}$ over 100~km away. The signal-to-background ratio of the single-site Heysham 2 signal represents the limit of what is practically achievable within a reasonable time frame for single-site anomaly detection and reactor measurement with a detector of this size. Scaling up the detector geometry to a larger size may enable anomaly detection and measurement of smaller reactors or more distant reactor signals, at standoffs around 200~km and beyond, in a complex nuclear landscape.

Gd-WbLS was found to be the most promising medium for reactor antineutrino detection. One of the limitations of  this medium is the loss of information due to the overlap of Cherenkov and scintillation light from an interaction in the detector. This could be mitigated using a slower scintillator~\cite{Biller2020} or a lower concentration of PPO~\cite{Onken2020}, preferably in combination with novel light collection such as fast photosensors~\cite{Adams2016} and wavelength-based photon sorting~\cite{Kaptanoglu2020}. Resulting improvements in particle identification \textit{e.g.}, using the isotropy of the Cherenkov light from particle interactions~\cite{BELLERIVE2016}, offer further potential to optimize sensitivity with this nascent technology.

For consistency with the single-event reconstruction, a 10-hit threshold has been used in the CoRe implementation for this study. However, with the improvement in reconstruction of IBD positrons at the lowest energies it is likely that a lower threshold is possible. CoRe and the subsequent Cobraa analysis combine to eliminate all or most of the backgrounds due to radioactivity. This, along with the ability to reconstruct well at lower energies, could bring the analysis energy threshold right down to the DAQ threshold. This is expected to increase the sensitivity of a detector and consequently bring down the dwell times output by the Cobraa analysis. Improvements in the reconstruction could benefit studies such as~\cite{Li2022}.

The machine-learning method has been used to reduce the fast neutron background. This could be extended to remove other sources of background by training other models to remove different event types. The models can then be run consecutively to extract all background sources by feeding forward the classified data from one model into the next. In this way, further background reduction in LEARN may be possible.

For the purposes of this study, both analysis pathways have assumed that we have no way to constrain the backgrounds through observation. In practice, it would be possible to make sideband analyses or background observations during reactor shutdowns. This has the potential to improve (\textit{i.e.}, reduce) dwell times, most notably through the reduction of systematic uncertainties in larger Gd-WbLS detectors. 

The two analysis pathways present a rate-only approach to remote reactor detection. They represent a first attempt to evaluate the potential of an antineutrino detector as a remote monitor. Making use of spectral information would provide additional information and offer the potential to improve sensitivities. A combination of the best of the two analysis pathways offers further potential to reduce the dwell times to a minimum. 

The toolchains presented in this study, or a combination of the two, could be used as a tool for non-proliferation to evaluate the sensitivity of any given detector to any given signal, with some assumptions about background rates and uncertainties on backgrounds. This could be useful for future prototypes, and particularly if antineutrino detectors are adopted for remote reactor monitoring in the field.

\begin{acknowledgments}
The authors would like to thank all those who worked on the adaptation of the RAT-PAC simulation to remote reactor monitoring for the WATCHMAN Scientific Collaboration including, but not limited to, M. Bergevin, M. Askins (IBD and fast neutron generators) and Z. Bagdasarian (WbLS model). Thanks also to F. Sutanto for development of the original analytical post-muon veto, which was adapted for this study. We would like to acknowledge M. Bergevin additionally for the development of the original reactor monitoring analysis, on which Cobraa in particular is based, and for the many useful conversations we had, as well as R. Foster for the initial studies of distant reactors.

This work was supported by the Atomic Weapons Establishment (AWE), as contracted by the Ministry of Defence, and the Science and Technology Facilities Council (STFC).
\end{acknowledgments}

\section*{Author contributions}

The Monte Carlo simulations were generated by L. Kneale and S. Wilson. 

L. Kneale developed the CoRe coincidence reconstruction, $\beta$-neutron rate and post-muon veto calculations, and the Cobraa analysis. Much of the material in Sections~\ref{sec:reactor-antinu}~to~\ref{sec:Cobraa} and Section~\ref{sec:muon-veto} was taken from the Ph.D. thesis by LK~\cite{Kneale2021}.

S. Wilson created the LEARN analysis chain, which incorporates contributions from Tara Appleyard (machine learning), James Armitage (likelihood ratio test) and Niamh Holland (machine learning).

S. Wilson drafted Section~\ref{sec:signalsims}, Section~\ref{sec:Learn}, and some of Section~\ref{sec:results}. L. Kneale drafted the remainder of the work, incorporating contributions from M. Malek and S. Wilson.

\providecommand{\noopsort}[1]{}\providecommand{\singleletter}[1]{#1}%
%

\end{document}